\PassOptionsToPackage{unicode=true}{hyperref}
\PassOptionsToPackage{naturalnames=true}{hyperref}
\documentclass[aps,rmp, %groupedaddress, 
superscriptaddress, reprint]{revtex4-2}
\usepackage{graphicx}
\usepackage[hidelinks]{hyperref}
\usepackage{bm}
\usepackage{amsmath}
\usepackage{amsthm}
\usepackage{amsfonts}
\usepackage{amssymb}
\usepackage{latexsym}
\usepackage{xcolor}
\usepackage[utf8]{inputenc}		
\usepackage{braket}
\usepackage{slashed}
\usepackage{siunitx}
\allowdisplaybreaks
\usepackage{longtable}
\usepackage{dcolumn}
\newcolumntype{.}{D{x}{}{-1}}
\newcolumntype{w}[1]{D{.}{.}{#1}}
\newcommand{\Za}{Z\alpha}

\newcommand{\nn}{\nonumber}

\makeatletter
\g@addto@macro\bfseries{\boldmath}
\makeatother
\def\lbar{\lambda\hskip-4.5pt\vrule height4.6pt depth-4.3pt width4pt}
\begin{document}
\preprint{Version 1.0}

\title{\texorpdfstring{Comprehensive theory of the  Lamb shift in light muonic atoms}{Comprehensive theory of the Lamb shift in light muonic atoms}}

\author{K. Pachucki}
\affiliation{\mbox{Faculty of Physics, University of Warsaw, Pasteura 5,  02-093 Warsaw,  Poland}}
\author{V. Lensky}
\affiliation{\mbox{Institut f\"ur Kernphysik, Johannes Gutenberg-Universit\"at Mainz, 55128 Mainz, Germany}}
\author{F. Hagelstein}
\affiliation{\mbox{Institut f\"ur Kernphysik, Johannes Gutenberg-Universit\"at Mainz, 55128 Mainz, Germany}}
\affiliation{\mbox{Paul Scherrer Institut, CH-5232 Villigen PSI, Switzerland}}
\author{S. S. Li Muli}
\affiliation{\mbox{Institut f\"ur Kernphysik, Johannes Gutenberg-Universit\"at Mainz, 55128 Mainz, Germany}}
\author{S. Bacca}
\affiliation{\mbox{Institut f\"ur Kernphysik, Johannes Gutenberg-Universit\"at Mainz, 55128 Mainz, Germany}}
\affiliation{\mbox{Helmholtz-Institut Mainz, Johannes Gutenberg-Universität Mainz, 55099 Mainz, Germany}}
\author{R. Pohl}
\affiliation{\mbox{Institut f\"ur Physik, Johannes Gutenberg-Universit\"at Mainz, 55099 Mainz, Germany}}

\date{\today}

\begin{abstract} 
A comprehensive theory of the  Lamb shift in light muonic atoms such as $\mu$H, $\mu$D, $\mu^3$He$^+$, and $\mu^4$He$^+$ is presented, 
with all  quantum electrodynamic corrections included at the precision level constrained by the uncertainty of  nuclear structure effects.
This analysis can be used in the global adjustment of fundamental constants and in the determination of nuclear charge radii.
Further improvements in the understanding of  electromagnetic interactions of light nuclei
will  allow for a promising test of  fundamental interactions by comparison with  ``normal" atomic spectroscopy, 
in particular, with  H-D and $^3$He-$^4$He isotope shifts.
\end{abstract}

\maketitle

\tableofcontents

\section{Introduction} 

Two-body systems such as hydrogen $(e^- p^+)$, positronium $(e^- e^+)$, and muonium $(e^- \mu^+)$
 have long been recognized as the best tools to verify  fundamental interaction theories \cite{Kinoshita:90}.
This is because their energy levels can be calculated analytically or numerically to a high precision, 
 limited in principle by the accuracy of fundamental physical constants.
 Starting with nonrelativistic quantum mechanics, the Hamiltonian of two charged particles with masses $m_1$ and $m_2$
 interacting with an attractive Coulomb potential, 
 \begin{align}
 H = \frac{\vec p_1^{\,2}}{2\,m_1} +   \frac{\vec p_2^{\,2}}{2\,m_2} -\frac{Z\,\alpha}{r} = E_k + H_0\,,
 \end{align}
 can be decomposed in terms of the total kinetic energy
 \begin{align}
 E_k = \frac{(\vec p_1+\vec p_2)^2}{2\,(m_1+m_2)}\,,
 \end{align}
 and the one-body Hamiltonian with the reduced mass $\mu$
 \begin{align}
 H_0 = \frac{\vec p^{\,2}}{2\,\mu} -\frac{Z\,\alpha}{r}\,, \label{10}
 \end{align}
 where 
 %$\vec q = (m_2\,\vec p_1-m_1\,\vec p_2)/(m_1+m_2)$. 
 $\vec p = -i\,\vec\nabla_{\! r}$ is the relative momentum of these two particles.
 For precise definition of constants and units, see Sec.~\ref{section2}. 
 The eigenvalues of this Hamiltonian
 \begin{align}
 E_{nl} = -\frac{(Z\,\alpha)^2\,\mu}{2\,n^2} \label{11}
 \end{align}
 depend on the principal quantum number $n=1,2,3,\ldots$ and not on the angular momentum number $l=0,1,\ldots, n-1$.
 The degeneracy of states with different $l$ is a characteristic feature of the nonrelativistic Coulomb Hamiltonian.
 We know, however, that a more accurate description of hydrogenlike levels must rely on the relativistic theory. 
 The first questions arise here:  What is the relativistic analog of the instantaneous Coulomb interaction, and what is the correct
 two-body equation for charged particles? In fact, there is no definitive answer to these questions yet.
 Only in the case in which the mass of one particle goes to infinity can we write a Dirac equation for the second spin-$1/2$ particle 
 (or a Klein-Gordon equation for a spin-$0$ particle) in the Coulomb potential of a static nucleus   \cite{Itzykson:1980rh}. 
 For a hydrogen atom having a proton mass that is approximately 2000 times larger than the electron mass, the Dirac equation is a good starting point. 
 It yields energy levels that depend on the principal quantum number $n$ (as in the nonrelativistic case) but also on the total angular momentum number $j$,
 as well as on the fine structure constant $\alpha$.
 Accordingly, the states $2S_{1/2}$ and $2P_{1/2}$ carry the same $j$ and are thus degenerate.
Here we use the historical notation, which is still used by atomic spectroscopists, where a state is labeled by its $n L_{j}$ 
with $S,P,D,F,\ldots$ standing for $l=0,1,2,3,\ldots$, and the subscript $j$ denoting the particular value of the total angular momentum. 
 However, in a true hydrogen atom the energy of the $2S_{1/2}$ state is slightly above that of the $2P_{1/2}$ one. 
 This splitting, first observed experimentally by \citet{Lamb:47} and subsequently named the Lamb shift,
was fundamental for the construction of quantum electrodynamics (QED)
by Feynman, Schwinger, and Tomonaga, for which they were awarded the Nobel Prize \cite{Dyson:65}.
 QED theory allows us to account in a perturbative manner for the finite nuclear mass \cite{Shabaev:98}, for the electron self-interaction
 and the vacuum polarization \cite{Itzykson:1980rh, Berestetskii:82}, and, in current use, for the accurate description of not only hydrogenlike but also of  arbitrary atomic systems \cite{Drake:23}. 
 All of these effects are described in Sec.~\ref{section3} for
such hydrogenlike systems, where the electron is replaced by the
 muon, a 200 times heavier lepton that is also a pointlike particle.

 In contrast to leptons, however, the nucleus in most cases cannot be treated as a pointlike particle.
 For example, a proton has a finite charge distribution that can be measured in lepton-proton scattering experiments, 
 as was first shown by \citet{PhysRev.103.1454}.
 At present the nuclear charge distribution cannot be calculated {\it ab initio}, 
 at least not with the accuracy needed by atomic spectroscopy measurements.
 This nuclear charge distribution affects the Coulomb interaction at small distances. Although it is a
small effect (about 1 MHz in H) it needs to be taken into account due to the
high accuracy of spectroscopic measurements; for example, the $1S-2S$ transition frequency in hydrogen \cite{Parthey:11} is
 \begin{align}
 \nu_\mathrm{H}(1S-2S) = 2\,466\,061\,413\,187\,035(10)\;\mathrm{Hz}\,.
 \end{align}
The finite proton size effect can in principle be determined from the comparison of 
 theoretical predictions for $\nu_\mathrm{H}(1S-2S)$ to the aforementioned measurement.
However, the Rydberg constant Ry enters into the comparison as a conversion
constant between experiment (measured in SI units) and theory (performed in atomic units)~\cite{Pohl:2016glp, codata18}.
Thus far the only  determination of Ry has been available from hydrogen itself 
because only this system can be calculated and measured accurately enough at the same time. 
 Therefore, we need a second transition like $2S-nS$ to determine the two unknowns \cite{codata18}: the Rydberg constant
 and the mean square proton charge radius; see Eq.\ (\ref{65}). Since the measurements of  $2S-nS$ and other transitions in hydrogen are much less accurate, the atomic spectroscopy determination of the proton charge radius
 was of limited accuracy: $r_p = 0.8768(69)$ fm \cite{Mohr:08}, consistent with the electron-proton scattering determination \cite{A1:2013fsc}.
 This situation was stable for a long time, until a new determination of
 the proton charge radius became available from the Lamb shift measurement in muonic hydrogen $\mu$H \cite{pohl2010}. 
 This new determination resulted in a much smaller
 proton charge radius of $r_p = 0.841\,84(67)$ fm and thus questioned the universality of electromagnetic interactions 
 and the validity of QED theory for composite particles \cite{Pohl:2013yb}.

This is why the comparison of nuclear charge radii obtained at first from $\mu$H \cite{pohl2010, antognini2013}, then from $\mu$D \cite{pohl2016}, 
$\mu^4$He \cite{krauth2021}, and $\mu^3$He \cite{crema:2023} to those obtained from ``normal" atomic spectroscopy
is a sensitive test of lepton universality and also a search for the existence of possible  yet unknown lepton-nucleus interactions 
at the scale from a few to a few hundred femtometers; these interactions have not yet been probed experimentally by other means~\cite{Pohl:2013yb,Carlson:2015jba}.
A similar or even stronger sensitivity to the lepton universality is expected from a direct comparison of the electron versus muon
scattering of the proton, which is the aim of the MUSE Collaboration \cite{Lorenzon:20206S}.
The charge radii of the proton and other light nuclei are also important for the determination of fundamental physical constants like 
the Rydberg constant from the spectroscopy of H \cite{codata18} or He$^+$ \cite{eikema, udem} and the electron-nucleus mass ratios from the spectroscopy of HD$^+$ \cite{hdplus1, hdplus2, hdplus3}.
In fact, the global adjustment of fundamental constants, performed
periodically every four years by CODATA \cite{codata18},
will now employ the nuclear charge radii obtained from muonic atom spectroscopy. 
Indeed, the most accurate determination of the root mean square (rms) nuclear charge radius $r_C$ is 
by the measurement of the $2S-2P$ transition in the hydrogenlike system, which consists of a muon and the nucleus \cite{pohl2010, antognini2013}. 
Owing to the 200-times-heavier muon, muonic atoms are much more
sensitive to the nuclear size and to nuclear structure effects than normal electronic atoms. In particular,
the rms radius shift of the muonic atom energy levels  is $\sim 200^3$ larger than that of the electronic ones.
Therefore, the determination of the nuclear charge radii from muonic atoms is much more accessible and precise.
For this purpose, one needs to calculate QED and nuclear structure effects on the energy levels 
accurately enough to be able to interpret the remainder as a finite nuclear size effect. 
\citet{Borie82} performed an extensive study of energy levels in muonic atoms by solving the Dirac equation 
with the muon mass replaced by the reduced mass of the muon-nucleus system, and by including the Breit interaction. 
This treatment partially accounts for the nuclear recoil corrections, but its results are not accurate enough for light muonic atoms.
Therefore, an approach suited to light atomic systems, exact in the mass ratio, was developed by \citet{muonic} 
and was widely followed in the later literature.

Here we present a comprehensive theory of the Lamb shift in light muonic atoms, with particular attention paid to the consistent separation of
a point-nucleus QED from the nuclear structure effects.  It is based mostly on the recent literature
[see reviews by \citet{muH_theory, muD_theory, mu3He_theory, mu4He_theory}, and references therein], with several contributions calculated  or recalculated here. 
All results are shown in Table~\ref{tab:recfns}, with each entry explained in its dedicated section. 
The crucial point is the preservation of consistency in the Lamb shift theory among all muonic and electronic atoms 
and, consequently, the consistent determination of nuclear charge radii. 
\begin{table*}[!ht]
\caption{Contributions to the $2P_{1/2}-2S_{1/2}$ energy difference $E_L$ in meV, with the charge radii $r_C$ given in fm. All corrections larger than 3\% of the overall uncertainty are included.
Theoretical predictions for $E_L$ are $E_L(\mathrm{theo}) = E_\mathrm{QED} + {\cal C}\,r_C^2 + E_\mathrm{NS}$. The last two rows show the values of $r_C$ determined from a comparison of $E_L(\mathrm{theo})$ to $E_L(\mathrm{exp})$.
\label{tab:recfns}}
\begin{ruledtabular}
%\scriptsize
\begin{tabular}{lllw{0.6}w{2.6}w{2.6}w{2.6}}
\multicolumn{1}{l}{Sec.}
& \multicolumn{1}{l}{Order}
& \multicolumn{1}{l}{Correction}
& \multicolumn{1}{c}{$\mu$H}
        & \multicolumn{1}{c}{$\mu$D}
        & \multicolumn{1}{c}{$\mu^3$He$^+$}
        & \multicolumn{1}{c}{$\mu^4$He$^+$}
  \\
\hline\\[-5pt]
\ref{s02} &  $\alpha\,(Z\alpha)^2$ &  eVP$^{(1)}$ & 205.007\,38 &  227.634\,70 & 1641.886\,2 & 1665.773\,1 \\
\ref{s02} & $\alpha^2(Z\alpha)^2$ &  eVP$^{(2)}$ & 1.658\,85 & 1.838\,04  & 13.084\,3 & 13.276\,9\\
\ref{s02} &   $\alpha^3(Z\alpha)^2$ &  eVP$^{(3)}$ & 0.007\,52 & 0.008\,42(7) &  0.073\,0(30)& 0.074\,0(30)\\
\ref{s06} &   $(Z, Z^2, Z^3)\,\alpha^5$ &  light-by-light eVP & -0.000\,89(2) & -0.000\,96(2)  & -0.013\,4(6) & -0.013\,6(6)\\
\ref{s01} & $(Z\alpha)^4$ &  recoil & 0.057\,47 &  0.067\,22 & 0.126\,5 & 0.295\,2\\
\ref{s03} &  $\alpha\,(Z\alpha)^4$ &  relativistic with eVP$^{(1)}$ & 0.018\,76 & 0.021\,78 & 0.509\,3 & 0.521\,1\\
\ref{s15} & $\alpha^2(Z\alpha)^4$ &  relativistic with eVP$^{(2)}$& 0.000\,17 & 0.000\,20 & 0.005\,6 & 0.005\,7\\
\ref{s09} &  $\alpha\,(Z\alpha)^4$ &$\mu$SE$^{(1)}$ + $\mu$VP$^{(1)}$,  LO  &-0.663\,45  &-0.769\,43   & -10.652\,5 &-10.926\,0 \\
\ref{s10} &  $\alpha\,(Z\alpha)^5$ &  $\mu$SE$^{(1)}$ + $\mu$VP$^{(1)}$,  NLO &-0.004\,43  & -0.005\,18& -0.174\,9 & -0.179\,7\\
\ref{a02} &  $\alpha^2(Z\alpha)^4$ &   $\mu$VP$^{(1)}$ with eVP$^{(1)}$ & 0.000\,13 & 0.000\,15 & 0.003\,8 & 0.003\,9 \\
\ref{s12} &  $\alpha^2(Z\alpha)^4$ &   $\mu$SE$^{(1)}$ with eVP$^{(1)}$ & -0.002\,54 & -0.003\,06  & -0.062\,7   & -0.064\,6\\
\ref{s07} & $(Z\alpha)^5$ &  recoil  & -0.044\,97  & - 0.026\,60  & -0.558\,1& - 0.433\,0\\
\ref{a01}      & $\alpha\,(Z\alpha)^5$ &  recoil with eVP$^{(1)}$  & 0.000\,14(14) & 0.000\,09(9) & 0.004\,9(49) & 0.003\,9(39) \\
\ref{s11} &  $Z^2\alpha\,(Z\alpha)^4$ &  nSE$^{(1)}$ &-0.009\,92  & -0.003\,10 & -0.084\,0&-0.050\,5 \\
\ref{s13} & $\alpha^2(Z\alpha)^4$ &  $\mu F_1^{(2)}$, $\mu F_2^{(2)}$, $\mu$VP$^{(2)}$& -0.001\,58 & -0.001\,84 & -0.031\,1 & -0.031\,9\\
\ref{a03}          & $(Z\alpha)^6$ &  pure recoil & 0.000\,09 & 0.000\,04 & 0.001\,9 & 0.001\,4\\
\ref{a04}          & $\alpha\,(Z\alpha)^5$ & radiative recoil & 0.000\,22 & 0.000\,13 & 0.002\,9 & 0.002\,3\\
\ref{s14} & $\alpha\,(Z\alpha)^4$ &  hVP & 0.011\,36(27)  & 0.013\,28(32) & 0.224\,1(53)& 0.230\,3(54) \\
\ref{a05} & $\alpha^2(Z\alpha)^4$ &  hVP with eVP$^{(1)}$& 0.000\,09 & 0.000\,10 &  0.002\,6(1) & 0.002\,7(1)\\[2ex]
\ref{f01} &  $(Z\alpha)^4$                &  $r_C^2$                             &-5.197\,5 \,r_p^2 & -6.073\,2\,r_d^2  & -102.523\,r_h^2 &-105.322\,r_\alpha^2\\
\ref{f02} &  $\alpha\,(Z\alpha)^4$    &  eVP$^{(1)}$ with $r_C^2$ &-0.028\,2 \,r_p^2  & -0.034\,0\,r_d^2  & -0.851\,r_h^2  & -0.878\,r_\alpha^2\\
\ref{f03} &  $\alpha^2(Z\alpha)^4$ &  eVP$^{(2)}$ with $r_C^2$ &-0.000\,2 \,r_p^2  & -0.000\,2\,r_d^2  & -0.009(1)\,r_h^2  & -0.009(1)\,r_\alpha^2 \\[2ex]
\ref{n01} &  $(Z\alpha)^5$ &  TPE     &0.029\,2(25)  & 1.979(20)  & 16.38(31)  & 9.76(40)\\
\ref{n02} &  $\alpha^2(Z\alpha)^4$ &  Coulomb distortion & 0.0 & -0.261 & -1.010 & -0.536\\
\ref{n03} & $(Z\alpha)^6$ &  3PE &-0.001\,3(3)& 0.002\,2(9) &-0.214(214)  &  -0.165(165) \\
\ref{n04} & $\alpha\,(Z\alpha)^5$ &  eVP$^{(1)}$ with TPE & 0.000\,6(1)    & 0.027\,5(4)  & 0.266(24)  & 0.158(12)\\
\ref{n05} & $\alpha\,(Z\alpha)^5$ & $\mu$SE$^{(1)}$ + $\mu$VP$^{(1)}$ with TPE &0.000\,4 & 0.002\,6(3) & 0.077(8) & 0.059(6)\\[2ex]
\ref{section3}  & $E_{\rm QED}$ &point nucleus & 206.034\,4(3) & 228.774\,0(3)& 1644.348(8) & 1668.491(7) \\
\ref{section4} & ${\cal C}\,r_C^2$ &finite size  &-5.225\,9\,r_p^2 & -6.107\,4\,r_d^2 & -103.383\,r_h^2 & -106.209\,r_\alpha^2 \\
\ref{section5}  & $E_{\rm NS}$ &nuclear structure  & 0.028\,9(25) & 1.750\,3(200)& 15.499(378) & 9.276(433) \\[2ex]
& $E_L$(exp) & experiment\footnote{Presented by
\citet{antognini2013}, \citet{pohl2016}, \citet{krauth2021}, and \citet{crema:2023}.\label{footnote1}}  & 202.370\,6(23) & 202.878\,5(34)& 1258.598(48)& 1378.521(48)\\[2ex]
& $r_C$ & this review & 0.840\,60(39) & 2.127\,58(78) & 1.970\,07(94) & 1.678\,6(12)\\
& $r_C$ & previous work$^\mathrm{a}$  & 0.840\,87(39) & 2.125\,62(78) & 1.970\,07(94)& 1.678\,24(83) \\
\end{tabular}
\end{ruledtabular}
\end{table*}

\section{Expansion of energy in powers of the fine structure constant $\alpha$}
\label{section2}
Throughout this review, we use the natural units $\hbar=c=1$. We start with the definition of the Lamb shift in the presence of the nuclear spin $\vec I$, the spin of the orbiting lepton $\vec S$, and the angular momentum $\vec L$.
The effective Hamiltonian in the subspace of states with a definite principal quantum number $n$, 
orbital momentum $l=0$ or $1$, and nuclear spin $I\leq 1$ is 
\begin{align}
 H_{\rm eff}(n,l) =&\  E_1 + E_2\, \vec S\cdot\vec L + E_3\,\vec S\cdot\vec I + E_4\,\vec L\cdot\vec I \nonumber \\ & \hspace{-11ex}
 + E_5\,(L^i\,L^k)^{(2)}\, (I^i\,I^k)^{(2)} + E_6\,(L^i\,L^k)^{(2)}\, I^i\,S^k\,,  \label{01}
\end{align}
where $(L^i\,L^k)^{(2)} = L^i\,L^k/2 + L^k\,L^i/2 -\vec L^2\,\delta^{ik}/3$. Here $i$ and $k$ are Cartesian indices; to distinguish them from Minkowski indices, the latter are denoted by lowercase greek letters. Furthermore, we use Einstein notation, which implies a sum over repeated indices. Note that we also limit the consideration to the case in which the orbiting lepton is a muon. Let $\vec J = \vec L+\vec S$; then
for the $S_{1/2}$ state $l=0,\ j=1/2$, and we define $E(nS_{1/2}) = E_1(n,0)$, while for the $P_{1/2}$ state
\begin{align}
E(nP_{1/2} )=&\ E_1(n,1)+E_2(n,1)\,\langle \vec S\cdot\vec L \rangle_{j=1/2} 
\nonumber \\ =&\ E_1(n,1)-E_2(n,1)\,,  \label{02}
\end{align}
so we calculate energies as if there were no nuclear spin couplings. Owing to the hyperfine mixing of 
the $P_{1/2}$ and $P_{3/2}$ states, this definition is not equivalent to the centroid energy but follows
the definitions assumed in the literature devoted to muonic atoms and those of CODATA~\cite{codata18}.

Having defined the Lamb shift 
\begin{align}
E_L = E(2P_{1/2}) - E(2S_{1/2})\,,  \label{03}
\end{align}
we employ an expansion in the fine structure constant $\alpha=e^2/(4\pi)$ with $e$ the proton charge,
to classify all important contributions and express $E_L$ as the sum of many terms that have a definite power
of $\alpha$ or $Z\alpha$ (where $Z$ is the nuclear charge in units of $e$) but may depend on the muon-nuclear mass ratio in a nontrivial way. For this we assume
that the electron vacuum polarization gives a single power of $\alpha$ (details are explained in Sec.~\ref{section3}). 
All corrections up to $\alpha^5$ are calculated with the exact mass dependence, while corrections of order $\alpha^6$
are obtained using the expansion in the muon-nucleus mass ratio up to the linear term only because these higher-order corrections are almost negligible.

To obtain the numerical values in Table~\ref{tab:recfns}, we use the following constants from the CODATA 2018 adjustment \cite{codata18}:
\begin{align}
\alpha^{-1} =&\ 137.035\,999\,084(21)\,,  \label{04} \\ 
m_\mu =&\; 105.658\,375\,5(23)\,\mbox{\rm MeV}\,,  \label{05} \\
\lbar_\mu =&\; 1.867\,594\,306(42)\,\mbox{\rm fm}\,,  \label{06}
\end{align}
where $m_\mu$ is the mass and $\lbar_\mu=1/m_\mu$ the reduced Compton wavelength of the muon. The conversion constant that connects the energy and length units is
\begin{align}
%197.3269804593024658
\hbar c =&\;  197.326\, 980\, 459\ldots \,\mbox{\rm MeV\,fm}\,.
\end{align}
The relevant mass ratios are
\begin{subequations}
\begin{align}
\frac{m_\mu}{m_e} =&\; 206.768\,283\,0(46) \,, \label{07a} \\    
\frac{m_\mu}{m_p} =&\; 0.112\,609\,526\,4(25)\,,   \label{07b} \\
\frac{m_\mu}{m_d} =&\; 0.056\,332\,718\,3(13)\,,   \label{07c} \\
\frac{m_\mu}{m_h} =&\; 0.037\,622\,379\,7(8)\,,   \label{07d} \\
\frac{m_\mu}{m_\alpha} =&\, 0.028\,346\,557\,7(6)\,, \label{07e}
\end{align}
\end{subequations}
where the subscripts $d,\ h,$ and $\alpha$ denote the deuteron, helion ($^3$He nucleus), and $\alpha$ particle ($^4$He nucleus), respectively. 
Moreover, with $\mu$ the reduced mass of the two-body system, 
\begin{align}
\mu = \frac{m_\mu}{1+m_\mu/M}\, \label{12}
\end{align}
with $M$ standing for the nuclear mass,
we define the ratio
\begin{equation}
\beta = \frac{m_e}{Z \alpha \, \mu} \,, \label{08}
\end{equation}
for which we obtain the following values:
\begin{subequations}
\begin{align}
\beta_p =&\  0.737\,383\,68\,, \label{09a} \\
\beta_d =&\  0.700\,086\,14\,, \label{09b} \\
\beta_h =&\  0.343\,842\,92\,, \label{09c}\\
\beta_\alpha =&\  0.340\,769\,14\,. \label{09d}
\end{align}
\end{subequations}
Finally, the nonrelativistic Coulomb wave function $\phi$ with nonrelativistic energy $E_0$
is the solution of $(H_0-E_0)\phi=0$, with $H_0$ as given by Eq. (\ref{10}) and $E_0 = E_{nl}$ from Eq. (\ref{11}).
The radial parts of the wave function for the states of interest are
\begin{align}
R_{20}(r) =&\ \frac{(\mu\,Z\,\alpha)^{3/2}}{\sqrt{2}}\,\exp\biggl(-\frac{\mu\,Z\,\alpha\,r}{2}\biggr)
\left(1-\frac{\mu\,Z\,\alpha\,r}{2}\right), \label{13} \\
R_{21}(r) = &\  \frac{(\mu\,Z\,\alpha)^{3/2}}{2\,\sqrt{6}}\,\exp\biggl(-\frac{\mu\,Z\,\alpha\,r}{2}\biggr)\,(\mu\,Z\,\alpha\,r)\,, \label{14}
\end{align}
and the wave function at the origin is
\begin{align}
\phi_{nl}^2(0) = \frac{R_{nl}^2(0)}{4\,\pi} = \frac{(\mu\,Z\,\alpha)^3}{\pi\,n^3}\,\delta_{l0}\,.\label{14.1}
\end{align} 
Note that our choice of electromagnetic units is specified by the definition of $\alpha$ in terms of $e$. However, the expressions for the relevant energies [Eqs.~\eqref{10} and \eqref{11}], the wave functions [Eqs.~\eqref{13}--\eqref{14.1}], and the final results for the energy shifts do not depend on this choice. At the same time, intermediate quantities such as the photon propagator may change if one uses different electromagnetic units.
\section{QED contributions to the Lamb shift}
\label{section3}
To calculate QED corrections to the energy levels, we assume at first that the nucleus is pointlike, while the nuclear size and nuclear structure
are considered separately in Secs.\ \ref{section4} and \ref{section5}.
A pointlike nucleus with spin $0$  satisfies the Klein-Gordon equation, a nucleus with spin $1/2$ satisfies the Dirac equation, and a nucleus with spin $1$ satisfies the Proca equation, 
with the last corresponding to a $g$~factor equal to $1$.
The radiative corrections on the nucleus line are included in the nuclear electromagnetic form factors and structure functions,
with an exception described in Sec.~\ref{s11}. 
As explained in Sec.~\ref{section2}, all corrections up to $\alpha^5\,m_\mu$ order are calculated with the exact muon-nuclear mass ratio, and
$\alpha^6\,m_\mu$ QED corrections are expanded in the mass ratio.
We now start with the leading QED effects. Since we specialize in the case of an orbiting muon, from this point on we suppress the label on the muon mass, denoting it as $m$ to make the equations more compact.

\subsection{Electron vacuum polarization}
\label{s02}

\begin{figure}[htb]
%\begin{center}
\includegraphics[scale=0.28]{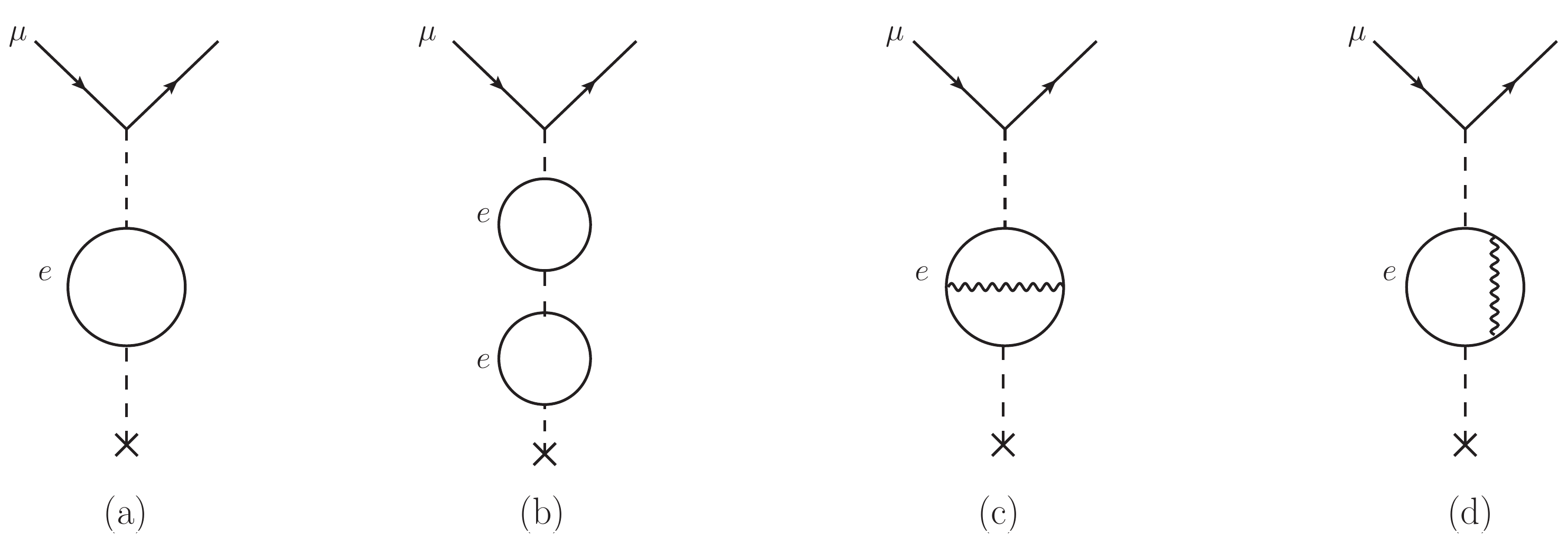}
\caption{Feynman diagrams for the pure QED electric vacuum polarization contribution to the Lamb shift. (a) Uehling potential. (b) K\"allen-Sabry potential, reducible two-loop part. (c), (d) K\"allen-Sabry potential, irreducible two-loop part.}
\label{FIG: eVP}
%\end{center}
\end{figure}

The electron vacuum polarization (eVP) (see Fig.~\ref{FIG: eVP}) modifies the photon propagator
\begin{equation}
-\frac{g^{\mu\nu}}{k^2}\rightarrow - 
\frac{g^{\mu\nu}}{k^2\,[1+\bar{\omega}(k^2/m_e^2)]}\,, \label{15}
\end{equation}
where $k^2 = (k^0)^2 -\vec k^{\,2}$ is the photon momentum squared.
The sum of one-particle irreducible diagrams $\bar{\omega}$ is expanded in a power series of $\alpha/\pi$
\begin{align}
\bar{\omega} = \bar{\omega}^{(1)} + \bar{\omega}^{(2)} + \bar{\omega}^{(3)} + \ldots\,, \label{16}
\end{align}
which results in the following expansion of the photon propagator:
\begin{align}
-\frac{g^{\mu\nu}}{k^2}\rightarrow - 
\frac{g^{\mu\nu}}{k^2}\,(1+\rho^{(1)} + \rho^{(2)} + \rho^{(3)} +\ldots)\,, \label{17}
\end{align}
where
\begin{align}
\rho^{(1)} =&\ -\bar{\omega}^{(1)}\,,  \label{18}\\
\rho^{(2)} =&\ -\bar{\omega}^{(2)} + (\bar{\omega}^{(1)})^2\,, \label{19}\\
\rho^{(3)} =&\ -\bar{\omega}^{(3)} + 2\,\bar{\omega}^{(1)} \,\bar{\omega}^{(2)} - (\bar{\omega}^{(1)})^3\,.  \label{20}
\end{align}
Each $\rho^{(i)}$ generates an eVP potential $V^{(i)}(r)$ at $k^0=0$
\begin{align}
V^{(i)}(r) = -Z\,\alpha\,\int \frac{d^3 k}{(2\,\pi)^3}\,\frac{4\,\pi}{\vec k^2}\,\rho^{(i)}(-\vec k^{\,2})\,e^{i\,\vec k\,\vec r}\,, \label{21}
\end{align}
and the corresponding corrections to the energy are
\begin{align}
E^{(1)} =&\ \langle V^{(1)}\rangle\,, \label{22} \\
E^{(2)} =&\ \langle V^{(2)}\rangle + \Bigl\langle V^{(1)}\, \frac{1}{(E_0-H_0)'}\,V^{(1)}\Bigr\rangle\,, \label{23} \\
E^{(3)} =&\ \langle V^{(3)}\rangle + 2\,\Bigl\langle V^{(2)}\, \frac{1}{(E_0-H_0)'}\,V^{(1)}\Bigr\rangle \label{24} \\
&\ \hspace*{-7ex}+\Bigl\langle V^{(1)}\, \frac{1}{(E_0-H_0)'}\,\big(V^{(1)} - \langle V^{(1)}\rangle\big)\, \frac{1}{(E_0-H_0)'}\, V^{(1)}\Bigr\rangle\,, \nonumber
\end{align}
where the prime in the denominator denotes a subtraction of the reference state.
For example,  at the one--loop level $V^{(1)}$ is
\begin{equation}
V^{(1)}(r)  =  -\frac{Z\,\alpha}{r}\,
\frac{\alpha}{\pi}\,\int_4^\infty \frac{d(\xi^2)}{\xi^2}\,e^{-m_e\,\xi\,r}\,u(\xi^2)\,, \label{25}
\end{equation}
where 
\begin{align}
u(\xi^2) =  \frac{1}{3}\sqrt{1-\frac{4}{\xi^2}}\,\left(1+\frac{2}{\xi^2}\right)\,, \label{26}
\end{align}
and similarly 
\begin{align}
\bar{\omega}^{(1)}(\zeta^2) = \frac{\alpha}{\pi}\,\zeta^2\,
\int_4^\infty\,d(\xi^2)\frac{1}{\xi^2(\xi^2-\zeta^2)}\,u(\xi^2)\,. \label{27}
\end{align}
Using the radial functions for the $2P$ and $2S$ states from Eqs.~\eqref{13} and \eqref{14}, we find that the one-loop vacuum polarization contribution
to the Lamb shift is
\begin{align}
E^{(1)}_L =&\ \mu\,(Z\,\alpha)^2\,\frac{\alpha}{\pi}\,
\int_4^\infty \frac{d(\xi^2)}{\xi^2}\,u(\xi^2)\, 
\frac{(\beta\,\xi)^2}{2\,(1+\beta\,\xi)^4} \,, \label{28}
\end{align}
with the numerical results presented in Table~\ref{tab:recfns}. All of these one\nobreakdash-, two-, and three-loop eVP contributions have already been obtained in the literature;
see \citet{sgk2013}  and references therein.
While the two-loop vacuum polarization (VP) is also known analytically \cite{twoloopvp}, the three-loop VP is known only numerically. 
It was first calculated for $\mu$H by \citet{kinoshita1999} and later corrected by \citet{sgkcorr1} as well as by \citet{kinoshita2009erratum}. 
For other muonic atoms, \citet{sgk2013}  obtained approximate values,
and the numerical values in Table~\ref{tab:recfns} are taken from Table I of their work.

\subsection{Light-by-light electron vacuum polarization}
\label{s06}

\begin{figure}[htb]
%\begin{center}
\includegraphics[scale=0.36]{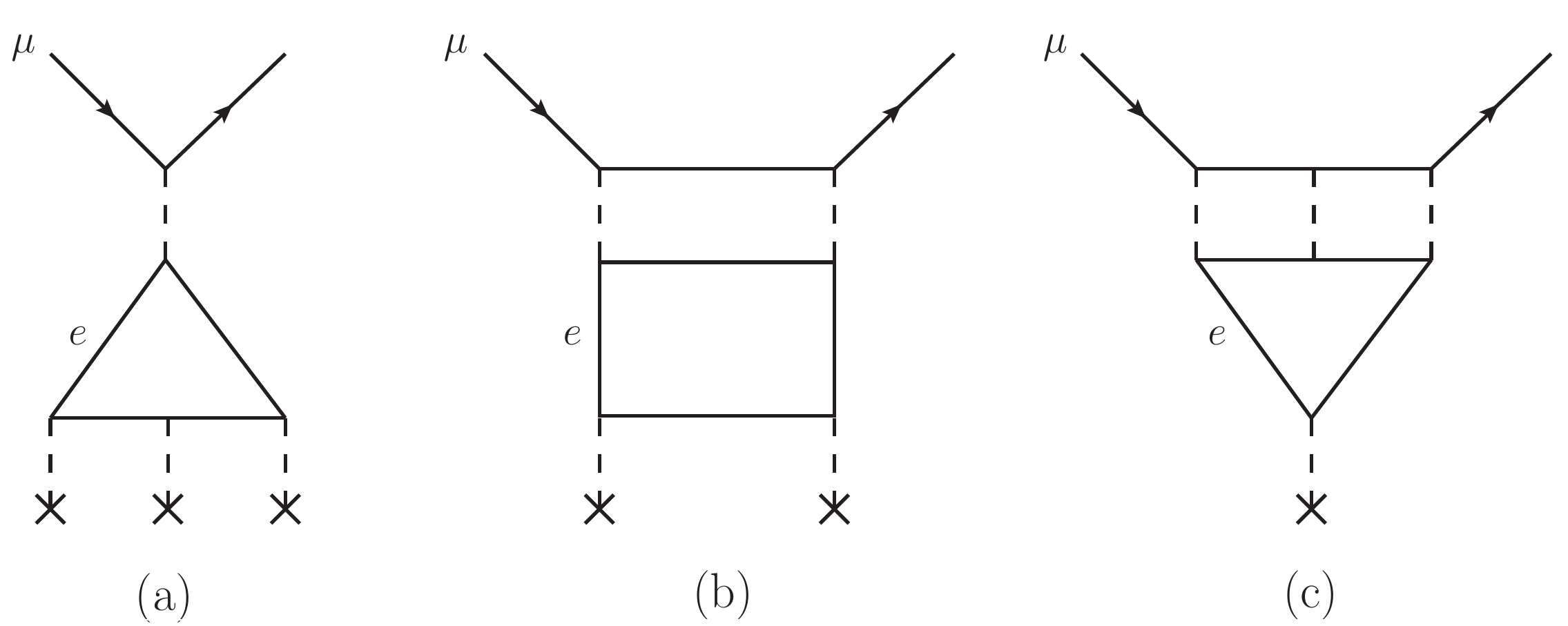}
\caption{Feynman diagrams for the light-by-light vacuum polarization contribution to the Lamb-shift. (a) Wichmann-Kroll correction. (b) Virtual Delbr\"uck scattering correction. (c) Inverted Wichmann-Kroll correction.}
\label{FIG: LbL}
%\end{center}
\end{figure}

This contribution comes from a closed electron loop with four photon legs. These legs can be attached in all possible ways to the muon and the nucleus lines; see Fig.~\ref{FIG: LbL}.
There are three types of diagrams with one, two, or three legs attached to the muon and the remaining legs to the nucleus.
Those with three legs attached to the nucleus are called the Wichmann-Kroll correction in the literature, those with two legs on each line 
are called the virtual Delbr\"uck scattering correction, and  those with one leg on the nucleus side we call here the inverted Wichmann-Kroll correction.
They were all calculated by \citet{Borie1978, sgk2010}, and \citet{sgk2013} (``LbL" in their Table I).
The overall  contribution is of the same order in $\alpha$ as the three-loop eVP but is about 10 times smaller.

\subsection{Leading recoil $\sim(Z\,\alpha)^4$}
\label{s01}
This is the leading-order nuclear recoil contribution. The nonrelativistic energies of the $2S$ and $2P$ states are the same, so the $(Z\,\alpha)^2$ recoil cancels out in the difference.
The leading $(Z\,\alpha)^4$ relativistic correction is almost the same; the difference is quadratic in the muon-nucleus mass ratio. 
It is derived starting with the expectation value of the Breit-Pauli Hamiltonian $H^{(4)}$ \cite{bs} with the nonrelativistic wave function, namely,
\begin{align}
\delta E =&\ \langle H^{(4)}\rangle\,,
\end{align}
where
\begin{align}
H^{(4)} =&\ -\frac{p^4}{8}\left(\frac{1}{m^3}+\frac{1}{M^3}\right)
-\frac{Z\,\alpha}{2\,m\,M}\,p^i\,
\biggl(\frac{\delta^{ij}}{r}+\frac{r^i\,r^j}{r^3}\biggr)\,p^j
\nonumber \\ &\
+\Bigl(\frac{1}{4\,m^2} +  \frac{1}{2\,m\,M}\Bigr)\,\frac{Z\,\alpha}{r^3}\, \vec r\times\vec p\cdot\vec \sigma
\nonumber \\ &\ 
+ \frac{\pi\,Z\,\alpha}{2}\,\Bigl( \frac{1}{m^2} + \frac{\delta_I}{M^2} + \frac{4}{3}\,r_{C}^2 \Bigr) \, \delta^{(3)}(\vec{r})\,, \label{29}
\end{align}
%\end{widetext}
and where $\delta_I=1$ for $I=1/2$, and $\delta_I=0$ for $I=0$ and $1$ by convention \cite{pachucki1995}. This results from the assumption
that the scalar particle satisfies the Klein-Gordon equation and the vector particle satisfies the Proca equation.
The Hamiltonian in Eq.~\eqref{29} includes the finite nuclear size correction, the treatment of which is deferred to Sec.~\ref{section4}; see Eq.~\eqref{64}. Without the finite size term, Eq.~\eqref{29} yields for the $2P_{1/2}-2S_{1/2}$ energy difference \cite{ulj2011a}
\begin{align}
\label{30}
\delta E_L =&  \left\{ \begin{array}{cc} 
\dfrac{(Z\alpha)^4 \mu^3}{48 \, M^2}  & \mbox{\rm for} \quad \delta_I = 1 \,, \\[2ex]
\dfrac{(Z\alpha)^4 \mu^3}{12 \, M^2}  & \mbox{\rm for} \quad \delta_I = 0 \,,
\end{array}  \right.
\end{align}
with the numerical results presented in Table~\ref{tab:recfns}. 

\subsection{Relativistic correction with the one-loop electron vacuum polarization}
\label{s03}
This is a contribution of the order $\alpha\,(Z\,\alpha)^4$  that combines the leading relativistic corrections with the one-loop eVP.
To derive it, we construct the photon propagator $G^{\mu\nu}$ in the modified Coulomb gauge.
What we mean is the following: We require the time component $G^{00}$ of the propagator to
coincide with the Coulomb potential including the vacuum polarization charge density, namely,
$G^{00} =\rho(\vec k^2)/\vec k^2$.
The transverse part of the propagator has to be of the form  \cite{pachucki2022}
\begin{align}
G^{ij}(k) =&\ \frac{\rho(-k^2)}{k^2}\,\biggl(\delta^{ij}-\frac{k^i\,k^j}{(k^0)^2}\biggr) - \frac{k^i\,k^j}{(k^0)^2}\,\frac{\rho(\vec k^2)}{\vec k^2}
\label{31}
\end{align}
in order to be equivalent  to the well-known propagator in the Feynman gauge
\begin{align}
G_F^{\mu\sigma}(k) = -\frac{g^{\mu\sigma}}{k^2}\,\rho(-k^2)\,.  \label{32}
\end{align}
For the evaluation of relativistic corrections with eVP, one needs the coordinate-space representation of the propagator
at $k^0=0$, which is
\begin{align}
G^{00}(\vec r) =&\  \int \frac{d^3k}{(2\pi)^3}\, e^{i\vec{k}\cdot\vec{r}}\,\frac{\rho({\vec k}^2)}{{\vec k}^2}\,,  \label{33}\\
G^{ij}(\vec{r})  =&\
-\frac{1}{2}\,\biggl(\delta^{ij} -\frac{r^ir^j}{r}\,\frac{d}{dr}\biggr)\,G^{00}(\vec r).  \label{34}
\end{align}
In the case of $\rho(\vec k^2)=1$ it becomes 
\begin{align}
G^{00}(\vec r) =&\ \frac{1}{4\,\pi\,r}\,,  \label{35}\\
G^{ij}(\vec{r})  =&\
-\frac{1}{8\,\pi}\,\biggl(\frac{\delta^{ij}}{r} +\frac{r^i\,r^j}{r^3}\biggr)\,,  \label{36}
\end{align}
the standard Coulomb gauge propagator at $k^0=0$.
One can now repeat the derivation of the Breit-Pauli Hamiltonian $H^{(4)}(V)$,
as done by \citet{vpbreit} using the aforementioned modified Coulomb propagator, 
\begin{align}
H^{(4)}(V) =&
-\frac{p^4}{8}\left(\frac{1}{m^3}+\frac{1}{M^3}\right)
+\frac{1}{8}\,\biggl(\frac{1}{m^2} + \frac{\delta_I}{M^2}\biggr)\,\nabla^2 V  \nonumber \\&
+\left(\frac{1}{4\,m^2}+\frac{1}{2\,m\,M}\right)
\frac{V'}{r}\,\vec L\cdot\vec\sigma \nonumber \\ &
+\frac{1}{2\,m\,M}\biggl[\nabla^2\left( V-\frac{1}{4}\,(r\,V)'\right) + \frac{V'}{r}\,\vec L^2
\nonumber \\ &
+\frac{p^2}{2}\,(V-r\,V')+(V-r\,V')\,\frac{p^2}{2}\biggr]\,,
\end{align}
and  obtain the correction 
\begin{align}
\delta E = \langle H^{(4)}(V^{(1)})\rangle + 2\,\biggl\langle V^{(1)}\,\frac{1}{(E_0-H_0)'}\,H^{(4)}\biggr\rangle\,,  \label{37}
\end{align}
where $H^{(4)} = H^{(4)}(-Z\,\alpha/r)$.
Equation \eqref{37} was first derived and calculated for $\mu$H by  \citet{muonic}, but with some mistakes.
We take the numerical values from Table I of \citet{ulj2011a},  who corrected these mistakes and calculated Eq.~(\ref{37}) for all nuclei of interest.
The use of $G^{ij}(\vec{r})$ from Eq.~(\ref{34}) will allow for future nonperturbative calculations of eVP corrections by solving the Schr\"odinger  
or Dirac equation numerically, which is much more efficient for heavier elements.

\subsection{Relativistic correction with the two-loop electron vacuum polarization}
\label{s15}
This correction is of the order $\alpha^2\,(Z\,\alpha)^4$ and can be obtained as in the one-loop case in Sec.~\ref{s03}. 
However, \citet{sgk2013} calculated it numerically using a slightly different approach that employed the Dirac equation.
The numerical values (see their Table VI)  are about $1\%$ of the one-loop case and are shown in Table~\ref{tab:recfns}.

\subsection{Leading muon self-energy and vacuum polarization}
\label{s09}
For the calculation of the one-loop muon self-energy $\mu$SE$^{(1)}$ and the muon vacuum polarization $\mu$VP$^{(1)}$
corrections to the Lamb shift, we rewrite the corresponding formula known for electronic hydrogen,
\begin{align} 
E(2S_{1/2}) = &\  \frac{1}{8}\,m\,\frac{\alpha}{\pi} \,(Z\,\alpha)^4\,
\left(\frac{\mu}{m}\right)^3\,
\biggl[\frac{10 }{9}-\frac{4}{15}
\nonumber \\ &\
-\frac{4}{3}\,\ln k_0(2S)+\frac{4}{3}\,\ln\left(
\frac{m}{\mu\,(Z\,\alpha)^2}\right) \biggr]\,,  \label{38}\\
E(2P_{1/2}) = &\  \frac{1}{8}\,m\,\frac{\alpha}{\pi} \,(Z\,\alpha)^4\!
\left(\frac{\mu}{m}\right)^3\!
\biggl[-\frac{1}{6}\,\frac{m}{\mu}-\frac{4}{3}\,\ln k_0(2P) \biggr], \label{39}
\end{align}
where $\ln k_0(n,l)$ is the Bethe logarithm,
\begin{eqnarray}
\ln k_0(2S) & = & 2.811\,769\,893\,1\ldots\,, \label{40} \\
\ln k_0(2P) & = & -0.030\,016\,708\,9\ldots\,, \label{41}
\end{eqnarray}
which is the same for electronic and for muonic hydrogenlike atoms.

\subsection{Next-to-leading muon self-energy and vacuum polarization }
\label{s10}
This is a two-photon exchange contribution accompanied by  the one-loop self-energy $\mu$SE$^{(1)}$ or vacuum polarization $\mu$VP$^{(1)}$.
For a point nucleus it is given by a contact interaction and thus has the same form for electronic and muonic hydrogenlike atoms, namely \cite{eides2001},
\begin{align}
\delta E(n,l) = \frac{\alpha\,(Z\,\alpha)^5}{\pi\,n^3}\,\frac{\mu^3}{m^2}\,4\,\pi\,\biggl(\frac{139}{128} + \frac{5}{192} - \frac{\ln 2}{2}  \biggr)\,\delta_{l0}\,, \label{42}
\end{align}
where the second term in parentheses comes from $\mu$VP$^{(1)}$.
There is a nuclear recoil correction to this formula that is considered in Sec.~\ref{a04},
and there is also a finite nuclear size correction that is considered in Sec.~\ref{n05}.

\subsection{Combined muon and electron vacuum polarizations}
\label{a02}
Correction to the energy due to  $\mu$VP$^{(1)}$ can be represented as a contact interaction,
 \begin{align}
\delta E =  -\frac{4}{15\,m^2}\,\alpha\,(Z\,\alpha)\,\langle \delta^{(3)}(r)\rangle =  -\frac{1}{15\,m^2}\,\frac{\alpha}{\pi}\,\langle \nabla^2 V\rangle, \label{43}
 \end{align}
 where $V=-Z\,\alpha/r$. Combining this contact interaction with the perturbation due to eVP$^{(1)}$, one obtains
\begin{align}
\delta E =  -\frac{2}{15\,m^2}\,\frac{\alpha}{\pi}\,\bigl[ \langle \nabla^2 V^{(1)}\rangle + 4\,\pi\,\Za\,\phi(0)\,\delta\phi(0)\bigr], \label{44}
\end{align}
where
\begin{align}
|\delta\phi\rangle = \frac{1}{(E_0-H_0)'} V^{(1)}|\phi\rangle\,. \label{45}
\end{align}
This correction was obtained by \citet{eides2001}, among many others. Particular values for the considered muonic atoms
were taken from \citet{sgk2013}.

\subsection{Muon self-energy combined with the electron vacuum polarization}
\label{s12}
This is similar to the previous correction, with  $\mu$VP$^{(1)}$ replaced by  $\mu$SE$^{(1)}$.
It is the one-loop muon self-energy in the Coulomb potential with one eVP$^{(1)}$ insertion.
For its derivation we generalize Eqs.\ \eqref{38} and \eqref{39} to an arbitrary potential 
\begin{align} 
\delta E = &\  \frac{\alpha}{4\,\pi\,m^2}\,\langle\phi|\nabla^2(V)|\phi\rangle\,
\biggl[\frac{10 }{9} +\frac{4}{3}\,\ln\left(\frac{m}{\mu\,(Z\,\alpha)^2}\right) \biggr]
\nonumber \\ & 
+\frac{2\,\alpha}{3\,\pi\,m^2}\,\langle\phi|\vec \nabla\,(H-E)\,\ln\Bigl(\frac{2\,(H-E)}{\mu\,(Z\,\alpha)^2}\Bigr)\,\vec \nabla\,|\phi\rangle
\nonumber \\ & 
+\frac{\alpha}{4\,\pi\,m\,\mu}\,\langle\phi|\frac{V'}{r}\,\vec L\cdot\vec\sigma|\phi\rangle\,,
\end{align}
where $H$ is the nonrelativistic Hamiltonian with the potential $V$ and eigenenergy $E$.
The perturbation due to $V^{(1)}$ was first estimated by \citet{muonic}. 
The complete calculation including the perturbed Bethe logarithm
was performed by \citet{ulj2011b} in their Eqs.~(29a)-(29d), and in Table~\ref{tab:recfns}
we use their results.

\subsection{Recoil $\sim(Z\,\alpha)^5$}
\label{s07}
This is the $(Z\,\alpha)^5$ contribution to the energy of two bound point particles, the muon and the nucleus,
without any radiative corrections. It vanishes in the limit of a heavy nucleus; therefore, we call it a recoil correction.
It  depends not only on the muon-nucleus mass ratio but also on  the value of the nuclear spin $I$.
The explicit formula  was derived originally for the spin $I=1/2$ nucleus by \textcite{Sa1952, Er1977}; this formula was valid for an arbitrary mass ratio.
Here we extend this formula to the case in which one of the particles has spin $I=0$ or $I=1$
using derivations presented in Sec.~\ref{n01}. The result is
\begin{align}
E(n,l) =&\ \frac{\mu^3}{m\,M}\,\frac{(Z\,\alpha)^5}{\pi\,n^3}
\biggl\{
\frac{2}{3}\,\delta_{l0}\,\ln\Bigl(\frac{1}{Z\,\alpha}\Bigr)  - \frac{8}{3}\,\ln k_0(n,l)
 \nonumber \\ &
-\frac{1}{9}\,\delta_{l0} 
-\frac{7}{3}\,a_n - 2\,\delta_{l0}\,\ln\Bigl(1+\frac{m}{M}\Bigr)\nonumber \\ &
+\frac{m^2}{M^2-m^2}\,\ln\biggl(\frac{M}{m}\biggr)\,\delta_{l0}\,[2+I\,(2\,I-1)]\biggr\},\label{46}
\end{align}
where
\begin{align}
a_n =&\ -2\left[\ln\Big(\frac{2}{n}\Big) + \Big(1+\frac{1}{2} + \ldots + \frac{1}{n}\Big)
+1-\frac{1}{2\,n}\right] \delta_{l0} \nonumber \\ &\ +
\frac{1-\delta_{l0}}{l\,(l+1)\,(2\,l+1)}\,. \label{47}
\end{align}
It agrees with that of \citet{shelyuto2018, shelyuto2019} for $I=0$ and $I=1$ nuclei under the assumption that $g=1$.
For other values of $g$, this recoil correction would have a logarithmic UV divergence.
Numerical results using Eq.~\eqref{46} for all nuclei are shown in Table~\ref{tab:recfns}.

\subsection{Recoil with the electron vacuum polarization}
\label{a01}
This is the  eVP$^{(1)}$ correction to the $(Z\,\alpha)^5$ contribution in Eq.~\eqref{46}.
It is quite difficult to calculate; in fact, it was obtained only by \citet{ulj2011b} and only in the logarithmic approximation.
The results shown in Table I are numerically small and suppressed with respect to the leading recoil correction given in Eq.~\eqref{46} by a factor of $\alpha$.
To account for nonlogarithmic terms, we assume a conservative 100\% uncertainty.

\subsection{Nuclear self-energy}
\label{s11}
%\begin{figure}[htb]
%%\begin{center}
%\includegraphics[scale=0.4]{Rad_T.pdf}
%\caption{Feynman diagrams for the radiative corrections to the forward Compton scattering off a nucleus.}
%\label{FIG: Radiative}
%%\end{center}
%\end{figure}
If we assume a pointlike nucleus with spin $1/2$, the contribution of the nuclear self--energy
for an arbitrary hydrogenic state is
\begin{align}
E(n,l) =&\ \frac{Z\,(Z\,\alpha)^5\,\mu^3}{\pi\,n^3\,M^2}\biggl[\left(
\frac{10}{9}+\frac{4}{3}\,\ln\frac{M}{\mu\,(Z\,\alpha)^2} \right)\,\delta_{l0}
\nonumber \\ &\ 
-\frac{4}{3}\, \ln k_0(n,l)\biggr]\,. \label{48}
\end{align}
For a nonpointlike nucleus there is a finite size correction.
The problem is that the nuclear  self-energy is modified by,
and modifies as well, the finite size effect. 
To incorporate the correction (\ref{48}) unambiguously, 
we must precisely specify the nuclear mean square charge radius.
The usual definition through the Sachs electric form factor 
\begin{equation}
\frac{r_C^2}{6} = \frac{\partial G_E({q}^2)}
{\partial ({q}^2)}\biggr|_{{q}^2 = 0} \label{49}
\end{equation}
is not correct at our precision level, because
$G_E$ cannot be uniquely defined in the presence of electromagnetic interactions. 
Following \citet{radrec} we propose a different definition using
the forward  scattering amplitude described by
\begin{equation}
T^{\mu\nu}(q) = -i\,\int d^4x\,e^{i\,q\,x}\,
\langle t|T\,j^{\mu}(x)\,j^{\nu}(0)|t\rangle\,,  \label{50}
\end{equation}
where $t=(M,0,0,0)$.
We consider the behavior of the dominant $T^{00}$ component at small $q^2$ and $p^2-M^2=(t+q)^2-M^2$.
For a pointlike particle without self-energy corrections, one finds that
\begin{align}
T^{00} =&\ \,\mbox{\rm Tr}\left[
\gamma^0\,\frac{1}{\slashed{p} -M}\gamma^0\,\frac{(\gamma^0+I)}{4}
\right] + (q\rightarrow-q)\nonumber \\ 
\approx&\ \frac{2\,M}{p^2-M^2}+ (q\rightarrow-q)\,. \label{51}
\end{align}
For a finite size particle without self-energy corrections
\begin{equation}
\gamma^\mu \rightarrow \Gamma^\mu = \gamma^\mu\,F_1(q^2) + i\frac{\sigma^{\mu\nu}}{2M}\,q_\nu\,F_2(q^2)\,, \label{52}
\end{equation}

\begin{figure}[tb]
%\begin{center}
\includegraphics[scale=0.4]{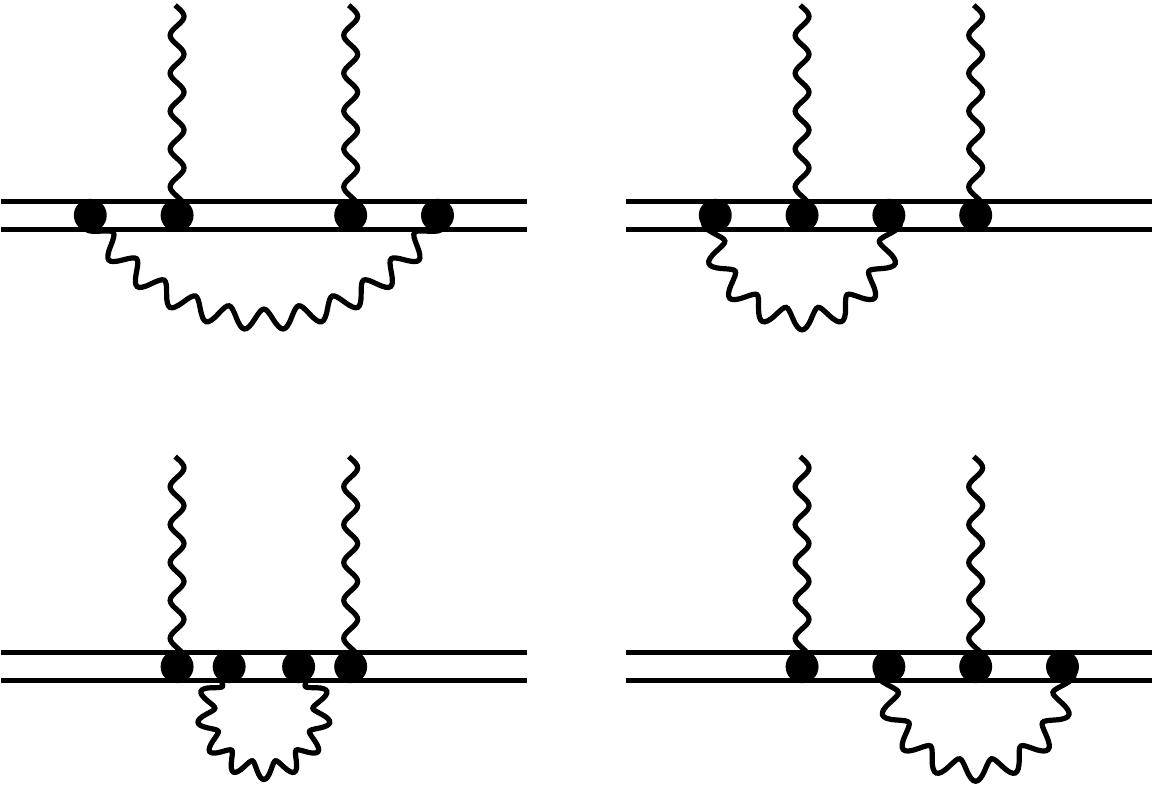}
\caption{Feynman diagrams for the radiative corrections to the forward Compton scattering off a nucleus.}
\label{FIG: Radiative}
%\end{center}
\end{figure}
$T^{00}$ acquires a correction
\begin{align}
\Delta T^{00} \approx&\ \frac{2\,M}{p^2-M^2}[G_E^2(q^2)-1] + (q\rightarrow-q)
\nonumber \\ \approx&\
\frac{2\,M}{p^2-M^2}\,q^2\,\frac{r_C^2}{3} + (q\rightarrow-q)\,, \label{53}
\end{align}
where $G_E = F_1 + \frac{q^2}{4\,M^2}\,F_2$.
The self-energy corrections for a pointlike particle coming from the diagrams in Fig.~\ref{FIG: Radiative} are \cite{radrec}
\begin{equation}
\Delta T^{00} = \frac{Z^2\,\alpha}{\pi\,M}\,\frac{q^2}{p^2-M^2}
\left(\frac{10}{9}+\frac{4}{3}\,\ln\frac{M^2}{M^2-p^2}\right)+ (q\rightarrow-q)\,. \label{54}
\end{equation}
We thus define $r_C^2$ using the following equation,
which describes the low-energy behavior of the correction to the
forward scattering amplitude of a pointlike particle:
\begin{align}
\Delta T^{00} =&\  \frac{q^2\,M}{p^2-M^2}\left(
\frac{4\,Z^2\alpha}{3\,\pi\,M^2}\,\ln\frac{M^2}{M^2-p^2}+
\frac{2}{3}\,r_C^2 \right)\nonumber \\ &\ + (q\rightarrow-q)\,.\label{55}
\end{align}
We expect that for any nucleus the aforementioned logarithmic term will be the same
because it is related only to the fact that the nucleus has a charge; 
it does not depend on other details like its spin. There is an arbitrariness
in the choice of the constant term, i.e., what belongs to the charge radius
and what belongs to the nuclear self-energy. The proposed definition separates
only the logarithmic term from the charge radius; thus,
the associated correction to the energy has the form
\begin{align}
E(n,l) =&\  \frac{2}{3\,n^3}\,(Z\,\alpha)^4\,\mu^3\,r_C^2\,\delta_{l0} 
\nonumber \\ &\hspace{-11ex}+
\frac{4\,Z\,(Z\,\alpha)^5}{3\,\pi\,n^3}\,\frac{\mu^3}{M^2}\,
\left[\ln\left(\frac{M}{\mu\,(Z\,\alpha)^2}\right)\delta_{l0}-\ln k_0(n,l)\right]\,, \label{56}
\end{align}
where the correction for $P$ states beyond $\ln k_0(n,l)$ goes into the nuclear magnetic moment.
The same formula for the nuclear self-energy will be assumed for all nuclei, and numerical results 
coming from the second line of Eq. (\ref{56}) are presented in Table~\ref{tab:recfns}.

\subsection{Muon two-loop form factors and vacuum polarization}
\label{s13}
This correction comes from the muon two-loop form factors and the two-loop vacuum polarization $\mu$VP$^{(2)}$:
\begin{align}
E(nS_{1/2}) =&\ \frac{\mu^3}{m^2}\,\Bigl(\frac{\alpha}{\pi}\Bigr)^2\,\frac{(Z\,\alpha)^4}{n^3}
\biggl( 4 F'_1(0) + F_2(0) -\frac{82}{81} \biggr)\,, \label{57}\\
E(nP_{1/2}) =&\ \frac{\mu^2}{m}\,\Bigl(\frac{\alpha}{\pi}\Bigr)^2\,\frac{(Z\,\alpha)^4}{n^3}\,\biggl( -\frac{1}{3}  \biggr)\,F_2(0)\,, \label{58}
\end{align}
where the muon two-loop form factors are \cite{barbieri1973}
\begin{align}
F'_1(0) =&\    -\frac{3 \zeta (3)}{4}-\frac{4819}{5184}-\frac{49 \pi ^2}{432}+\frac{1}{2} \pi ^2 \ln 2 
\nonumber \\ &\hspace{-9ex} 
+ \biggl[ \frac{1}{9}\,\ln^2\frac{m}{m_e}
-\frac{29}{108}\, \ln\frac{m}{m_e}+\frac{\pi^2}{54}+\frac{395}{1296}+ O\Bigl(\frac{m_e}{m}\Bigr)\biggr] \label{59}
\end{align}
and
\begin{align}
F_2(0) =&\     \frac{3 \zeta (3)}{4}+\frac{197}{144}+\frac{\pi ^2}{12}-\frac{1}{2} \pi ^2\ln 2 \nonumber \\
               &\ + \biggl[\frac{1}{3}\,\ln\frac{m}{m_e}-\frac{25}{36}+ O\Bigl(\frac{m_e}{m}\Bigr)\biggr]\,. \label{60}
\end{align}
The terms in square brackets in Eqs.~\eqref{59} and \eqref{60} come from the closed electron loop and thus are dominant.
Numerical results for all muonic atoms of interest are presented in Table~\ref{tab:recfns}.

\subsection{Pure recoil $\sim (Z\,\alpha)^6$}
\label{a03}
The $(Z\,\alpha)^6$ contribution to the energy of a bound system of two particles is expanded in the mass ratio $m/M$.
The nonrecoil term coincides with the Dirac energy and thus vanishes in the $2P_{1/2} - 2S_{1/2}$ difference.
The leading term is linear in the mass ratio and is given by \cite{recoil1, recoil2}
\begin{align}
\delta E_L = -\frac{m^2}{M}\, \frac{(Z\,\alpha)^6}{8}\, \biggl(\frac{1}{3}  + 4\,\ln 2 - \frac{7}{2}\biggr), \label{61}
\end{align}
which results in a relatively small correction; see Table~\ref{tab:recfns}.

\subsection{Radiative recoil $\sim\alpha\,(Z\,\alpha)^5$}
\label{a04}
The $\alpha\,(Z\,\alpha)^5$ contribution to the energy is given by a contact interaction and is thus proportional to $\phi^2(0)$.
We expand the coefficient in powers of the muon-nucleus mass ratio $m/M$. The nonrecoil term was already accounted for in Sec.~\ref{s10};
the next term in the mass ratio expansion is the radiative recoil correction \cite{radrec, eides2001},
\begin{align}
\delta E_L = \frac{\mu^3}{m\,M}\, \frac{\alpha\,(Z\,\alpha)^5}{8}\,1.364\,49\,, \label{62}
\end{align}
which includes $\mu$SE$^{(1)}$ and $\mu$VP$^{(1)}$.

\subsection{Hadronic vacuum polarization}
\label{s14}
To estimate the effect of the hadronic vacuum polarization (hVP), we assume ``the most realistic value" according to \citet{sgk2021} (``Scatter" in their Table 4).
Using as a reference the energy shift due to  $\mu$VP$^{(1)}$ [the second term in Eq.~\eqref{38}], we write the hVP contribution as
\begin{align}
E(n,l) =&\ \frac{\mu^3}{m^2}\,\frac{\alpha}{\pi} \,\frac{(Z\,\alpha)^4}{n^3}\,\biggl(-\frac{4}{15} \biggr)\,\gamma_{\rm had}\,\delta_{l0}\,. \label{63}
\end{align}
Equation \eqref{63} differs from the $\mu\text{VP}^{(1)}$ term by a factor of $\gamma_{\rm had} = 0.6746(160)$, thus giving an appreciable effect that should be included in the same way in muonic and electronic atoms to obtain consistent nuclear charge radii. The corresponding numerical values are shown in Table~\ref{tab:recfns}.

\subsection{Combined electron and hadronic vacuum polarization}
\label{a05}
We represent this correction as the aforementioned coefficient $\gamma_{\rm had}$ times the correction due to $\mu$VP$^{(1)}$ combined with eVP$^{(1)}$ from Sec.~\ref{a02}.

\section{Finite nuclear size contribution}
\label{section4}
All corrections in this section are proportional to the mean square charge radius and thus have the form ${\cal C}\,r_C^2$.

\subsection{Leading finite size $r_C^2$}
\label{f01}
The definition of the rms charge radius $r_C^2$ depends on the nuclear spin
and, in particular, there are different definitions for a spin 1 particle,
such as the deuteron, as discussed by \citet{Jentschura2011}. For a particle with spin $I$ and mass $M$, $r_C^2$ can be  defined through the effective interaction 
with the electromagnetic field,
\begin{eqnarray}
\delta H &=& e\,A^0  -
e\,\biggl(\frac{r_C^2}{6}+\frac{\delta_I}{8\,M^2}\biggr)\,
\vec\nabla\cdot\vec E
\nonumber \\ &&
-\frac{e}{2}\,\frac{Q}{I\,(2\,I-1)}\,(I^i\,I^j)^{(2)}\,\nabla^j E^i-\frac{\mu_I}{I}\,\vec I\cdot\vec B\,,\quad
\label{64}
\end{eqnarray}
where $\mu_I$ and $Q$ are the magnetic dipole and electric quadrupole moments, and the Darwin-Foldy term $\delta_I$ 
has been defined after Eq.\ (\ref{29}). Namely, for a scalar particle $\delta_0 = 0$, and for a spin-$1/2$ particle the Dirac equation gives $\delta_{1/2} = 1$.
For a vector particle, we assume that the charge radius is defined with respect to the Proca particle,
namely, the point vector particle with $g=1$ and $Q=0$, and this gives $\delta_1=0$ \cite{pachucki1995}.
This convention coincides with the definition employed in nuclear physics \cite{filin2021}, and affects the relativistic recoil correction (see Sec.~\ref{s01}), 
while the finite nuclear size correction is
\begin{equation}
E_{\rm FNS}(n,l) = \frac{2\,\pi}{3}\,Z\,\alpha\,\phi^2(0)\,r_C^2 = \frac{2}{3\,n^3}\,(\Za)^4\,\mu^3\,r_C^2\,\delta_{l0}. \label{65}
\end{equation}
Apart from the spin dependence, the nuclear self-energy affects the definition of $r_C$.
This is described in Sec.~\ref{s11}, where following \citet{radrec} we propose using the forward two-photon exchange amplitude
for the precise definition of the nuclear charge radius.  

\subsection{One-loop electron vacuum polarization  with $r_C^2$}
\label{f02}

\begin{figure}[htb]
%\begin{center}
\includegraphics[scale=0.4]{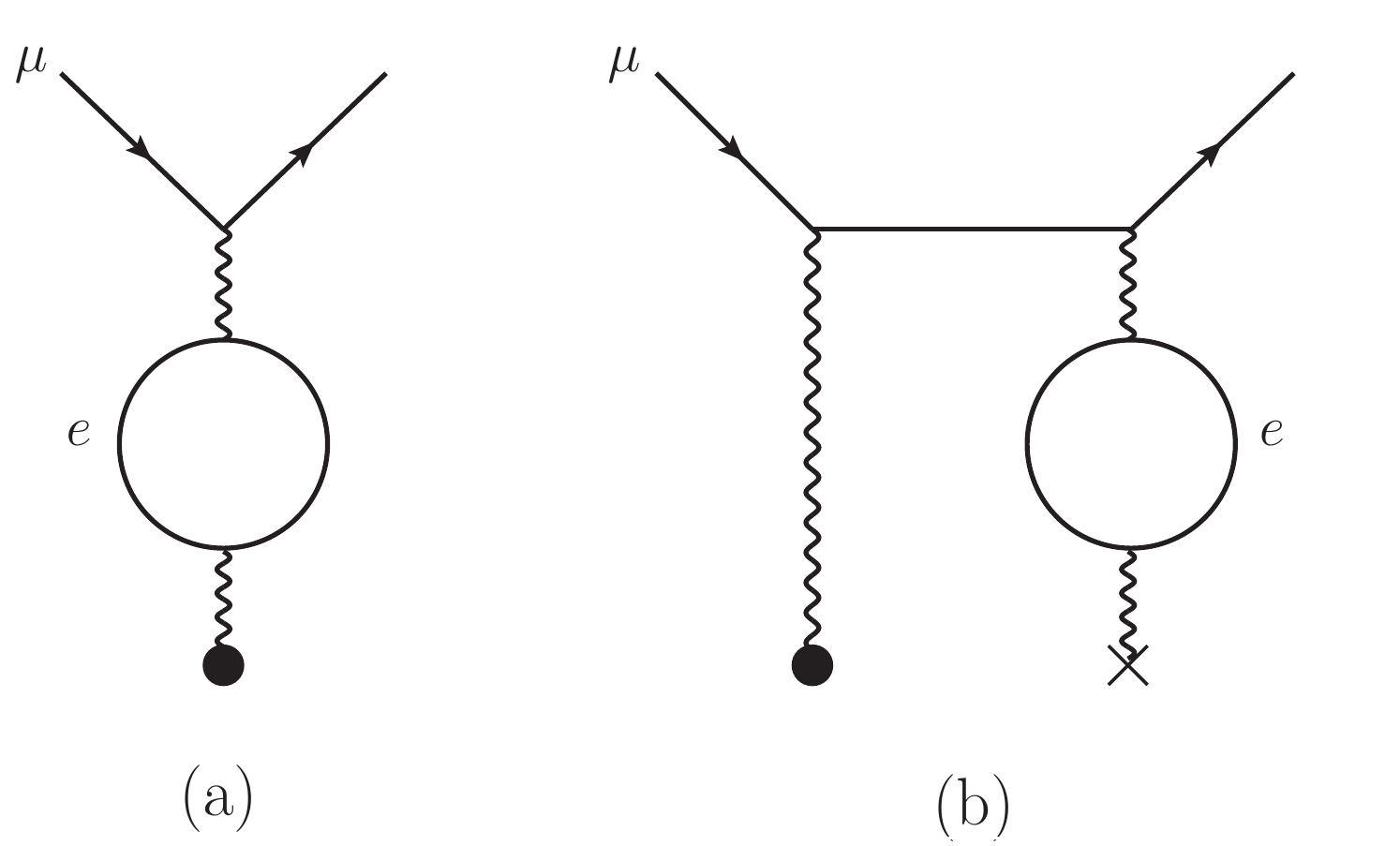}
\caption{One-loop electron vacuum polarization corrections to the nuclear finite size. (a) Photon propagator correction. (b) Wave function correction.}
\label{FIG: FS_eVP}
%\end{center}
\end{figure}

The leading QED correction to the finite size contribution
is due to the one-loop eVP and is described by two terms \cite{muonic}, corresponding to the two diagrams in Fig.~\ref{FIG: FS_eVP},
\begin{align}
\delta E_{\rm FNS} = \frac{r_C^2}{6}\bigl[\langle \nabla^2 V^{(1)}\rangle +  8\,\pi\,\Za\,\phi(0)\,\delta\phi(0)\bigr]. \label{66}
\end{align}
The correction  is proportional to $r_C^2$, and the coefficient is presented in Table~\ref{tab:recfns}.

\subsection{Two-loop electron vacuum polarization with $r_C^2$}
\label{f03}
This is a correction similar to the previous one
but is  suppressed by an additional factor of $\alpha$. Thus, it is almost negligible.
It was calculated by \citet{martynenko2014} for $\mu$D in their Eqs.~(30)--(32), 
and by \citet{krutov2015} for $\mu$He$^+$ (items 18 and 19 in their Table I).
Because they neglected third-order perturbation theory diagrams, we have added  a conservative uncertainty of 10\% to their results. 
Finally, the result for $\mu$H was obtained by rescaling it from $\mu$D.

\section{Nuclear structure contributions}
\label{section5}
The nuclear structure contributions beyond the finite nuclear size are expanded in powers of the fine structure constant $\alpha$, as with all other corrections.
We call the leading term of the order of $(Z\,\alpha)^5$ the two-photon exchange (TPE).
There are several corrections of higher order in $\alpha$, which  are all considered in separate sections.
Moreover, we assume that the possible radiative corrections on the nucleus line are all included in  $E_{\rm TPE}$,
with the exception of the leading nuclear self-energy considered in Sec.~\ref{s11}.

\subsection{Two-photon exchange}
\label{n01}
The $(Z\,\alpha)^5$ TPE contribution in a muonic atom with a nucleus of spin $I$ is given by \cite{munucl}
\begin{align}
E_\mathrm{TPE} =& -\frac{(Z\,e^2)^2}{2}\, \phi^2(0)\,\int_s\frac{d^4q}{(2\,\pi)^4\,i}\,
\frac{1}{q^4}\nonumber \\ & \times
\Bigl[T^{\mu\nu}-t^{\mu\nu}(I,M)\Bigr]\,t_{\mu\nu}(m)
\nonumber \\ =& 
-2\,(Z\,e^2)^2\,\phi^2(0)\,\frac{m}{M}\,\int_s\frac{d^4q}{(2\,\pi)^4\,i}\nonumber \\ &\hspace*{-5ex} \times
\frac{[T_2-t_2(I,M)](q^2-\nu^2)-[T_1-t_1(I,M)]\,(q^2+2\,\nu^2)}
{q^4\,(q^4-4\,m^2\nu^2)}\,, \label{67}
\end{align}
where $T^{\mu\nu}$ is the forward virtual Compton scattering amplitude, defined in Eq.\ (\ref{50}), 
that can be expressed in terms of two Lorentz invariant functions $T_1(\nu,-q^2)$ and $T_2(\nu,-q^2)$
\begin{align}
T^{\mu\nu} =&
-\biggl(g^{\mu\nu}-\frac{q^\mu\,q^\nu}{q^2}\biggr)\,\frac{T_1}{M}
\nonumber \\ &\ 
+\biggl(\frac{t^\mu}{M}-\frac{\nu}{q^2}\,q^\mu\biggr)\,
\biggl(\frac{t^\nu}{M}-\frac{\nu}{q^2}\,q^\nu\biggr)\,\frac{T_2}{M} \label{68}
\end{align}
and where $\nu = q^0$  is the lab-frame photon energy. 
To be consistent with the $(Z\,\alpha)^5$ recoil correction in Eq.\ (\ref{46}),  
we assume in Eq.~(\ref{67}) that $t^{\mu\nu}(I,M)$ corresponds to the pointlike nucleus of spin $I$. 
For $I=1/2$, $t^{\mu\nu}(M) \equiv t^{\mu\nu}(1/2, M)$, and
\begin{align}
t^{\mu\nu}(M) = {\rm Tr}\biggl[\gamma^\mu\frac{1}{\slashed{p}-M}\,
\gamma^\nu\,\frac{\gamma^0+I}{4}\biggr]+(q\rightarrow -q)\,, \label{69}
\end{align}
with $p = t+q$. From Eq.~\eqref{69} one obtains for a point Dirac particle
\begin{eqnarray}
t_1(1/2,M) &=& -\frac{4\,M^2\,\nu^2}{q^4-4\,M^2\,\nu^2}\,, \label{70} \\
t_2(1/2,M) &=& \frac{4\,M^2\,q^2}{q^4-4\,M^2\,\nu^2}\,. \label{71}
\end{eqnarray}
For a point scalar particle one obtains
\begin{eqnarray}
t_1(0,M) &=& 1\,, \label{72} \\
t_2(0,M) &=& \frac{4\,M^2\,q^2}{q^4-4\,M^2\,\nu^2}\,, \label{73}
\end{eqnarray}
and for a Proca vector particle \cite{lee1962}
\begin{eqnarray}
t_1(1,M) &=& -\frac{2\,\nu^2\,(6\,M^2 - q^2) - q^4}{3\,(q^4-4\,M^2\,\nu^2)}\,, \label{74} \\
t_2(1,M) &=& \frac{2\,q^2(6\,M^2\ - q^2)}{3\,(q^4-4\,M^2\,\nu^2 )}\,. \label{75}
\end{eqnarray}
The subscript $s$ in the integral in Eq.\ (\ref{67}) denotes an additional subtraction of a $1/q^5$ singularity, which 
has to be  proportional to $r_C^2$, provided the subtraction of a point nucleus with an appropriate spin is assumed.

We now make a digression regarding the pure recoil $(Z\,\alpha)^5$ correction.
It was originally calculated for the point spin-$1/2$ nucleus.
The difference between an arbitrary spin $I$ and a spin-$1/2$  point nuclei of the same mass is given by
\begin{align}
\delta E =&
-2\,(Z\,e^2)^2\,\phi^2(0)\,\frac{m}{M}\,\int_s\frac{d^4q}{(2\,\pi)^4\,i}\nonumber \\ & \hspace{-7ex}\times
\frac{[t_2(I)-t_2(1/2)](q^2-\nu^2)-[t_1(I)-t_1(1/2)]\,(q^2+2\,\nu^2)}
{q^4\,(q^4-4\,m^2\nu^2)}\,, \label{76}
\end{align}
using the aforementioned $t_{1,2}$ functions  with the mass argument implicit. This $\delta E$ gives the term proportional to $I\,(2\,I-1)$ in Eq.\ (\ref{46}), which generalizes
the pure recoil correction to the case of spin $I=0,1$ point nuclei, while for higher spins this integral diverges.

We now make a second digression.
$T^{\mu\nu}$  is a complete forward virtual Compton scattering amplitude and thus includes radiative corrections. 
Consequently, it has a $\ln(M^2 - p^2)$ singularity at the threshold; see Eq.\ (\ref{55}) with $p=t+q$ and $t=(M,\vec 0)$.
This singularity comes from the nuclear self-energy, and thus the corresponding $\ln q/q^5$  singularity should also be subtracted out in Eq.~(\ref{76}). 
The last subtraction is not mentioned in any calculation of the TPE correction from the scattering amplitudes, but it should
because $T^{\mu\nu}$ on the nucleus line is a complete amplitude. 
Moreover, the presence of the logarithmic singularity at threshold indicates the lack of the possibility to separate $T^{\mu\nu}$ into elastic and inelastic contributions.
However, we neglect this singularity and disregard the associated difficulties in what follows because the related effect is negligible at the current precision level.

\subsubsection{TPE in $\mu$H}
Returning to the calculation of TPE in Eq. (\ref{67}), we find that it  is conventionally split into a Born and a polarizability part,
\begin{align}
E_{\rm TPE}(\mu{\rm H}) =&\ E_{\rm Born} + E_{\rm pol}\,. \label{77}
\end{align}
The Born contribution
\begin{align}
E_{\rm Born} =&\ E_{\rm Fri} + E_{\rm rec} \label{78}
\end{align}
 in the infinite nuclear mass limit is given by 
\begin{align}
E_{\rm Fri} =&\ -\frac{\pi}{3}\,\phi^2(0)\,(Z\,\alpha)^2\, \mu\,r^3_\mathrm{F}\,, \label{79}
\end{align} 
where $r_\mathrm{F}$ is the Friar radius
\begin{align}
r^3_\mathrm{F} =&\ \int d^3r_1\int d^3r_2\,\rho(r_1)\,\rho(r_2)\,|\vec r_1-\vec r_2|^3\,, \label{80}
\end{align}
and the remainder is given by the small recoil correction $E_\mathrm{rec}$.
The presence of $\mu$ instead of $m$ in Eq.\ (\ref{79}) is a matter of convention, and this affects the definition of the TPE recoil correction.
For the polarizability part 
\begin{align}
E_{\rm pol} =&\ E_{\rm sub} + E_{\rm inel}, \label{81}
\end{align}
one can use dispersion relations to express $T_1$ and $T_2$ in terms of proton structure functions measured 
in electron-proton scattering. In the case of $T_1$, a once-subtracted dispersion relation is needed, 
giving rise to the subtraction function $T_1(0,-q^2)$, which cannot be measured directly, 
but has to be modeled or predicted from chiral perturbation theory ($\chi$PT), covariant \cite{Alarcon:2013cba,lensky2018}, or heavy baryon \cite{birse2012,Peset:2015zga,Peset:2014yha,Peset:2014jxa}.
The inelastic structure functions needed for the dispersive evaluation are known only
for the proton, deuteron, and helion,
although not in the entire kinematic region, especially for the deuteron and helion;
thus, a different approach will have to be employed for nuclei other than the proton.

In the case of the $2S$ state of $\mu$H, the Friar contribution
\begin{align}
E_{\rm Fri}= -0.021\,1(2)\; \mbox{\rm  meV} \label{82}
\end{align}
is obtained using the recent Friar radius  $r^3_\mathrm{F}(p) = 2.310(26)$ fm$^3$ from \citet{lin2021}.
Their work is a dispersive fit of nucleon form factors based on data in both the spacelike and timelike regions. 
Note that the dispersive analyses of the proton form factor \cite{Mergell:1995, Belushkin:2006qa, Lorenz:2014yda} 
predicted the smaller proton charge radius $r_p = 0.84$~fm before the $\mu$H Lamb shift measurement. These analyses have improved since then, taking into consideration new data from, for example,  the Mainz \cite{mainz2010} and Jefferson Lab (JLab) \cite{JLab2020} electron-proton scattering measurements. For a recent review of the history and the theoretical framework of the dispersive form factor analyses, see \citet{Lin:2021umz}.

The recoil correction was considered by \citet{sgk2015}, who obtained
\begin{align}
E_{\rm rec} = 0.000\,03(5)\; \mbox{\rm  meV}. \label{83}
\end{align}
A similar result recently obtained by  \citet{tomalak2022} 
\begin{align}
E_{\rm rec} = 0.000\,05(1)\; \mbox{\rm  meV}, \label{83t}
\end{align}
was based on the proton form factor parametrization of \citet{Borah:2020gte},
who used the small proton charge radius from the $\mu$H Lamb shift as a constraint. 

In total the Born contribution amounts to 
\begin{equation} 
E_\mathrm{Born}=-0.021\,1(2)\,\mathrm{meV}. \label{EBorn}
\end{equation}
This prediction not only is more precise but also differs from older values:  $E_\mathrm{Born}=-0.018\,6(16)\,\mathrm{meV}$ \cite{tomalak2019} 
and $E_\mathrm{Born}=-0.024\,7(16)\,\mathrm{meV}$ \cite{birse2012}. 
To explain this, we note that \textcite{tomalak2019, birse2012} used proton form factor parametrizations that corresponded to a large $r_p$. 
Since the value of the latter is correlated with the resulting $E_\mathrm{Born}$, a consistent TPE evaluation should 
use a form factor parametrization that results in a small $r_p$, as argued by \citet{Karshenboim:2014maa}.  
While \citet{tomalak2019}  used a procedure suggested by \textcite{Karshenboim:2014maa,sgk2015} to correct for the large radius 
of the A1 parametrization \cite{mainz2010,A1:2013fsc}, the comparison to Eq.~(\ref{EBorn}) indicates 
that the suggested correction might not be sufficiently accurate. 

The  subtraction contribution was considered in various works \cite{birse2012, Gorchtein:2013yga, Peset:2014jxa, tomalak2016, lensky2018},
and we take the value  
\begin{align}
E_{\rm sub} = 0.004\,6(24)\; \mbox{\rm  meV}\, \label{84}
\end{align}
from the most recent prediction in the framework of $\chi$PT \cite{lensky2018}.
As prevously mentioned, the $q^2$ dependence of the $T_1(0,-q^2)$ subtraction function, which is related to the magnetic dipole polarizability,
has not been experimentally constrained, and its uncertainty is the largest among all contributions to the $\mu$H Lamb shift.
Last, for the inelastic contribution 
\begin{align}
E_{\rm inel} = -0.012\,7(5)\; \mbox{\rm meV}\, \label{85}
\end{align}
we take a value of \citet{carlson2011}. As a final result for the TPE with the subtracted point proton, we obtain
\begin{align}
E_{\rm TPE}(\mu{\rm H},2S) =-0.029\,2(25)\; \mbox{\rm  meV}\,, \label{86}
\end{align}
with the uncertainty dominated by the one from the subtraction term. Note that, given the present uncertainties of the Friar and polarizability contributions, the recoil correction is negligible in the case of $\mu$H.

\subsubsection{TPE in $\mu$D}
In the case of $\mu$D, the first data-driven dispersive evaluation of the TPE correction was performed by \citet{carlson2014} with the result
$E_{\rm TPE}(\mu {\rm D},2S) = -2.011(740)$ meV, but with an inconsistent subtraction of the point deuteron $t^{\mu\nu}$. 
They set $G_C=1$ and $G_M=G_Q=0$ for a point deuteron,
whereas they should have set $G_C=G_M=1$ and $G_Q=0$, which would correspond to $G_1=G_2=1$ and $G_3=0$ (in their notation) because the $(Z\,\alpha)^5$ 
recoil correction in Eq.\ \eqref{46} was calculated assuming these values for the elastic form factors of the point deuteron. 
The same formalism as was used by \citet{carlson2014} was employed in several more recent works \cite{acharya2021, lensky2022a}. 
To make it consistent with Eq.~(\ref{46}), 
it is sufficient to modify the elastic contribution of \citet{carlson2014}, as shown in Appendix~\ref{app:subtraction}. 
The numerical effect of this modification turns out to be small ($\sim -0.000\,04$~meV) and thus can be neglected.

A remark regarding the dispersion relation formalism is in order. The subtraction function $T_1(0,-q^2)$ is treated differently in composite nuclei than in $\mu$H. In a data-driven approach, the dominant purely nuclear response in electron-nucleus scattering can be separated from the response of the individual nucleons, leading to a finite-energy sum rule for the nuclear part of $T_1(0,-q^2)$, as shown by \citet{Gorchtein:2015eoa}. As an alternative to using data, one often utilizes the nuclear response functions calculated from a theory of nuclear interactions. In this case, there is also as a rule no need for a subtraction, at least when the theory does not yet resolve the structure of the individual nucleons. That said, the small subtraction contribution due to the individual nucleons inherits all of the difficulties of the $\mu$H case.

Most recent works use chiral effective field theories ($\chi$EFT) of nuclear interactions \cite{epelbaum2009, machleidt2011, hammer2020, epelbaum2020}
to evaluate the deuteron structure functions instead of using
experimental input, due to the lack of quality data; see the discussion given by \citet{acharya2021}. \citet{lensky2022a, lensky2022b} analyzed pionless effective field theory ($\slashed{\pi}$EFT) and $\chi$EFT predictions and 
pointed out that the deuteron charge form factor parametrization by the JLab $t_{20}$ Collaboration \cite{JLABt20:2000qyq} does not describe 
the deuteron well enough in the low $q^2$ region, i.e., in the region without data. 
The latter parametrization was employed by \textcite{carlson2014} and \citet{acharya2021} to evaluate the elastic TPE.
Accordingly, we update the elastic part of \citet{acharya2021} by taking the value $-0.4456(18)$ meV from Table II of \citet{lensky2022a},
a result stemming from a $\chi$EFT calculation of $G_C$ \cite{filin2021}.
For the inelastic part  of \citet{acharya2021} we take the arithmetic mean of their two results  $-(1.511 + 1.519)/2$ meV
and finally add their hadronic contribution $-0.028$ meV to obtain
\begin{align}
E_{\rm TPE}(\mu{\rm D}, 2S) = &\ [-0.446(2) -1.515(15) -0.028(2)]\;\mbox{\rm meV} \nonumber \\ = &\  -1.990(15)\; \mbox{\rm  meV}\,. \label{87}
\end{align}
The most recent calculation by \citet{lensky2022a, lensky2022b} used $\slashed{\pi}$EFT amplitudes
for the forward virtual Compton scattering off the deuteron \cite{lensky2021} to obtain a similar sum of three contributions
\begin{align}
E_{\rm TPE}(\mu{\rm D}, 2S) = &\ [-0.446(8) -1.509(16) -0.032(6)]\;\mbox{\rm meV}  \nonumber \\ = &\  -1.987(20)\; \mbox{\rm meV}\,,\label{89}
\end{align}
which is in perfect agreement with Eq.~(\ref{87}).
Another recent calculation~\cite{Emmons:2020aov} used $\slashed{\pi}$EFT with pointlike nucleons to evaluate the deuteron inelastic structure functions.
This calculation obtained  the inelastic part of $E_\mathrm{TPE}$ [$-1.574(80)$~meV] with a larger central value (as a result of treating the nucleons as pointlike)
but also a larger uncertainty, making it consistent with the other evaluations.

Regarding the direct calculation of $E_{\rm TPE}$ from the nuclear theory, one can use  an effective Hamiltonian, 
either phenomenological or rooted in $\chi$EFT, and derive the TPE correction. 
The first such calculations were performed by \citet{leidemann1995}.
Later,  a much improved method was introduced in \citet{pachucki2011} and expanded on by \textcite{hernandez2014, pachucki2015}, and \citet{bacca2018},
resulting in the following formula for the TPE contribution:
\begin{align}
  E_{\rm TPE} =&\ E_{\rm nucl1} + E_{\rm nucl2} + E_{\rm pol}+\ldots\,, \label{90}\\
  E_{\rm nucl1} =&\ -\frac{\pi}{3}\,m\,\alpha^2\phi^2(0)
  \Big[Z\,R_{F}^3(p) + (A-Z)\,R_{F}^3(n) \Bigr]\,,\label{91} \\
  E_{\rm nucl2} =&\ -\frac{\pi}{3}\,m\,\alpha^2\phi^2(0)
  \sum_{i,j=1}^Z\langle\phi_N||\vec r_i-\vec r_j|^3|\phi_N\rangle\,,\label{92} \\
E_{\rm pol} =&\   
-\frac{4\,\pi\,\alpha^2}{3}\,\phi^2(0)\int_{E_T}dE\,
\sqrt{\frac{2\,\mu}{E}}\,|\langle\phi_N|\,\vec{\!d}\,|E\rangle|^2, \label{93}
\end{align}
where the Coulomb distortion correction is considered separately in Sec.~\ref{n02} since it is of $(Z\,\alpha)^6$ order.
Here $E_{\rm nucl1}$ is a sum of TPE contributions from each individual nucleon, $E_{\rm nucl2}$ is due to TPE
with different nucleons, and $E_{\rm pol}$ is the leading nuclear polarizability correction originating from the low-energy TPE
and is given by the matrix elements of the electric dipole operator $e\,\vec d$ between the nuclear ground state 
$|\phi_N\rangle$ and excited states $|E\rangle$, with $E_T$ the lowest excitation energy. 
Note that Eq.\ (\ref{91}) is proportional to $m$ instead of $\mu$, thus differing from  the convention in Eq.\ (\ref{79}).
The parameters $R_{F}(p)$ and $R_{F}(n)$ are the effective proton and neutron radii, 
which include the complete TPE with the corresponding nucleon.
The value for the proton is obtained from $E_{\rm TPE}$ in $\mu$H,
\begin{align}
R^3_{F}(p) = 2.876(246)\; {\rm fm}^3, \label{rfp}
\end{align}
and for the neutron the value was calculated by \citet{tomalak2019},
\begin{align}
R^3_{F}(n) = 0.712(223)\; {\rm fm}^3. \label{rfn}
\end{align}
There are many more small corrections in the effective Hamiltonian approach for the calculation of  the TPE contribution in $\mu$D; they are denoted by dots in Eq.\ (\ref{90}), and were separately calculated  by two groups: (i) \citet{pachucki2011} and \citet{pachucki2015} and (ii) \citet{hernandez2014, hernandez2017, hernandez2019} and \citet{bacca2018}. 
Using the aforementioned values of $R_{F}(p)$ and $R_{F}(n)$, we update the calculation of \citet{pachucki2015} by changing the single-nucleon contributions [see Eqs.~(45) and (46) in their paper] to $\delta_P E = -0.034(3)$ meV and $\delta_N E = -0.008(3)$ meV. Their total $E_\mathrm{TPE}$ thus becomes
\begin{align}
E_{\rm TPE}(\mu{\rm D}, 2S) = -1.961(20)\; \mbox{\rm meV}. \label{94}
\end{align}
Considering more elaborate calculations by the second group \cite{bacca2018}, we note that the point deuteron $(Z\,\alpha)^5$ recoil correction was not properly subtracted 
but do not expect this to be significant. Therefore, we take their $\delta^A_{\rm TPE} = [-1.675(15) - 0.262]$ meV (with the subtracted Coulomb distortion)
and add the nucleon contribution $\delta^N_{\rm TPE} = \delta_P E + \delta_N E$
to obtain
\begin{align}
E_{\rm TPE}(\mu{\rm D}, 2S) =&\  -1.979(15)\; \mbox{\rm meV}, \label{95}
\end{align}
which is in agreement with the updated value in Eq.\ (\ref{94}).
As a final value we take the mean value of Eqs.\ \eqref{87}, \eqref{89}, \eqref{94}, and \eqref{95}
and keep the largest uncertainty
\begin{align}
E_{\rm TPE}(\mu{\rm D}, 2S) = -1.979(20)\; \mbox{\rm meV}\label{96}
\end{align}
to account for possible systematic uncertainties in all of these determinations; see the discussion in Sec.~\ref{section6}.

\subsubsection{TPE in $\mu$He$^+$}
A calculation of $E_{\rm TPE}$ in $\mu ^3$He$^+$ using the experimentally measured inelastic structure functions
was performed by \citet{carlson2017}, but with
large uncertainties.
Much greater accuracy is achieved by direct nuclear structure calculations using Eqs.\ \eqref{90}--\eqref{93}.
In the case of $\mu^3$He$^+$ and $\mu^4$He$^+$, calculations of $E_{\rm TPE}$ were performed by \citet{chen2013, bacca2018} and \citet{dinur2016}.
We start with their results, which were denoted by $\delta_{\rm TPE}$ by \citet{bacca2018} in their Table 7, subtract the Coulomb distortion corrections
$\delta^{(0)}_C(\mu ^3{\rm He}^+) = 1.010$ meV and $\delta^{(0)}_C(\mu ^4{\rm He}^+) = 0.536$ meV, and update the single-nucleon contributions using Eqs.~(\ref{rfp}) and~(\ref{rfn}): 
$\delta^N_{\rm Zem} + \delta^N_{\rm pol} = -0.647(55)$ meV for $\mu ^3$He$^+$ and $\delta^N_{\rm Zem} + \delta^N_{\rm pol} = -0.738(63)$ meV 
for $\mu ^4$He$^+$. This gives 
\begin{align}
E_{\rm TPE}(\mu ^3{\rm He}^+, 2S)  =& 
%[-15.49(33) -1.010(x) + 0.52(3) \nonumber \\ & + 0.25(13)  - 0.647(55)]\;\mbox{\rm meV}  \nonumber \\ =& 
-16.38(31)\;\mbox{\rm meV}, \label{97} \\
E_{\rm TPE}(\mu ^4{\rm He}^+, 2S)  =& %[- 9.37(44) - 0.536(x) + 0.54(3) \nonumber \\ & + 0.34(20) - 0.738(63)]\;\mbox{\rm meV}\nonumber \\ =& 
-9.76(40)\;\mbox{\rm  meV},\label{98}
\end{align}
where the improvement in the accuracy with respect to the original results of \citet{bacca2018} is due to the updated single-nucleon contributions.
%These results are included in Table I.

\subsection{Coulomb distortion correction}
\label{n02}
Among corrections of higher order in $\alpha$, 
the largest one is the Coulomb distortion correction, which comes from the expansion in the ratio of the muon binding energy 
to the nuclear excitation energy and is enhanced by a factor of $\ln(Z\,\alpha)^2$,
\begin{align}
\delta_{C} E =& \frac{Z^4\,\alpha^6\,\mu^4}{6}\!
\int_{E_T}\frac{dE}{E}
\biggl[\frac{1}{6}+\ln\biggl(\frac{2\,\mu\,(Z\,\alpha)^2}{E}\biggr)\biggr]
\nonumber \\ &\times
|\langle\phi_N|\,\vec{\!d}\,|E\rangle|^2\,.\label{99} 
\end{align}
Since the first term $1/6$ in square brackets is much smaller than the logarithm, it is sometimes neglected \cite{bacca2018}.
This correction was calculated in a manner similar to the leading nuclear polarizability correction in Eq.\ (\ref{93}) [see \citet{bacca2018} for details],
with the numerical values presented in Table~\ref{tab:recfns}.

\subsection{Three-photon exchange}
\label{n03}
In the nonrecoil limit, the elastic three-photon exchange (3PE) contribution can be obtained by solving the Dirac equation with a finite size nucleus.
The corresponding relative $O(\alpha^2)$ correction to the finite size effect can be represented as \cite{pachucki2018}
\begin{align}
E^{(6)}_{\rm FNS}(2S) =&-(Z\,\alpha)^6\,m^3\,r_C^2\,\frac{1}{12}\biggl[\ln(m\,r_{C2}\,Z\,\alpha)+ \gamma_E -\frac{31}{16}\biggr]
        \nonumber \\
        & + (Z\,\alpha)^6\,m^5\,r_C^4\,\frac{1}{18}\,\biggl[\ln(m\,r_{C1}\,Z\,\alpha) + \gamma_E +  \frac{5}{2} \biggr] 
        \nonumber \\ &        
        +(Z\,\alpha)^6\,m^5\,r_{CC}^4\,\frac{1}{480}  \label{101}\,,
\end{align}
\begin{align}                
E^{(6)}_{\rm FNS}(2P_{1/2}) =&\ (Z\,\alpha)^6\,m\,\biggl(\frac{m^2\,r_C^2}{64} + \frac{m^4\,r_{CC}^4}{480}\biggr)\,, \label{102}
 \end{align}
where $r_{CC}^4 = \langle r^4\rangle$ is the fourth moment of the charge density, the effective nuclear charge radii $r_{C1}$ and $ r_{C2}$
encode the high-momentum contributions, and $\gamma_E$ is the Euler constant. For the exponential distribution of the nuclear charge $r_{CC}/r_C = 1.257\,433$,
$r_{C1}/r_C= 1.090\,044$, and $r_{C2}/r_C = 1.068\,497$. The inelastic contribution is difficult to estimate.
The only known result is for $\mu$D \cite{pachucki2018}, where it was found to partially cancel the elastic part.
Therefore, for other composite nuclei we assume that the total 3PE in the nonrecoil limit
is just a half of the elastic part, with the uncertainty being the other half.

\subsection{Electron vacuum polarization with TPE}
\label{n04}
The eVP$^{(1)}$  correction to the TPE contribution is
\begin{align}
\delta E_\mathrm{TPE} =& -\frac{(Z\,e^2)^2}{2}\, \phi^2(0)\,\int_s\frac{d^4q}{(2\,\pi)^4\,i}\,
\frac{1}{q^4}\,\Bigl[T^{\mu\nu}-t^{\mu\nu}(I,M)\Bigr]\nonumber \\ &\ 
\times t_{\mu\nu}(m)\,\biggl[- 2\,\bar\omega^{(1)}(q^2/m_e^2) + 2\,\frac{\delta\phi(0)}{\phi(0)}\biggr]\,, \label{103}
\end{align}
where $\delta\phi$ is as defined in Eq.~(\ref{45}), $\bar\omega^{(1)}$ is as defined in Eq.~(\ref{27}), and the subscript $s$ in the integral
denotes an additional subtraction of the finite size that is the same as in Eq. (\ref{67}).
The eVP$^{(1)}$ correction to the elastic part in the nonrecoil limit, namely, to the Friar term, was
obtained by \citet{sgk2018}. Their results were $0.000\,4,\; 0.007\,1,\; 0.212$, and $0.139$ meV
for the Lamb shifts in $\mu$H, $\mu$D, $\mu^3$He$^+$, and $\mu^4$He$^+$, respectively.
However, these results are incomplete because the eVP$^{(1)}$ correction to the inelastic part is unknown. 
Therefore, we do not use these results, and instead calculate the eVP$^{(1)}$ corrections to 
the leading contributions to $E_\mathrm{TPE}$, namely, $E_\mathrm{nucl1}$, 
$E_\mathrm{nucl2}$, and $E_\mathrm{pol}$ in Eqs.\ \eqref{90}--\eqref{93}. The details are presented in Appendices \ref{AppendixEVP}, \ref{AppendixEVA}, and \ref{AppendixEpol}, respectively. This largely completes the evaluation
of the eVP$^{(1)}$ correction with TPE.

In particular, for $\mu$H there is only the eVP$^{(1)}$ correction with TPE on the proton, which amounts to 
\begin{align}
\delta E_\mathrm{TPE}(\mu\mathrm{H}) =  0.000\,6(1)\; \mathrm{meV}
\end{align}
for the Lamb shift; see Eq. (\ref{B5}).
For $\mu$D, the resulting correction to the Lamb shift is the sum of the single-nucleon eVP$^{(1)}$ 
with TPE from Eq. (\ref{B5}) and the eVP$^{(1)}$ polarizability correction from \citet{kalinowski2019}, 
\begin{align}
\delta E_\mathrm{TPE}(\mu\mathrm{D}) =  0.027\,5(4)\; \mathrm{meV},
\end{align}
for the Lamb shift.
This is in agreement with the calculation of  the eVP$^{(1)}$ correction to TPE recently performed by \citet{lensky2022b, lensky2022a}, which gave $0.0274(3)$ meV.
Finally, for $\mu^3$He$^+$ and $\mu^4$He$^+$  we add $\delta E_\mathrm{nucl1}$ from Eq. (\ref{B5}),
$\delta E_\mathrm{nucl2}$ from Eq. (\ref{C11}), and $\delta E_\mathrm{pol}$ from Eq. (\ref{D3}) to obtain
\begin{align}
\delta E_\mathrm{TPE}(\mu^3\mathrm{He}^+) =&\ 0.266(24)\;\mathrm{meV}\,,\\
\delta E_\mathrm{TPE}(\mu^4\mathrm{He}^+) =&\ 0.158(12)\;\mathrm{meV}
\end{align}
for the Lamb shift. These results are shown in Table~\ref{tab:recfns}.

\newpage
 
\subsection{Muon self-energy and vacuum polarization with TPE}
\label{n05}
This correction is given by 
\begin{align}
\delta E_\mathrm{TPE}  =& -\frac{(Z\,e^2)^2}{2}\phi^2(0)\!\int\!\frac{d^4q}{(2\,\pi)^4\,i}
\frac{1}{q^4}\Bigl[T^{\mu\nu}-t^{\mu\nu}(I,M)\Bigr]\nonumber \\ &\ 
\times\bigl[t^{\rm rad}_{\mu\nu}(m) - 2\,\bar\omega^{(1)}(q^2/m^2)\,t_{\mu\nu}(m)\bigr]\,,\label{104}
\end{align}
where $t^{\rm rad}_{\mu\nu}$ is the muon self-energy correction to the pointlike  $t_{\mu\nu}$
and $\bar\omega^{(1)}$ is the one-loop muon vacuum polarization; see Eq.\ (\ref{27}).
An analytic expression for $t^{\rm rad}_{\mu\nu}$ is known \cite{radrec}, so this correction can in principle be calculated
for $\mu$H and $\mu$D using the experimentally or theoretically determined nuclear amplitudes $T_1$ and $T_2$.
For $\mu$He$^+$, the most convenient approach is the direct calculation using the nuclear theory with 
an effective Hamiltonian; cf.\ Eq.\ (\ref{90}). However, such a calculation has not yet been performed. The correction in Eq.\ (\ref{104}) has thus far been calculated \cite{sgk2018, faustov2017} in the elastic limit for an infinitely heavy nucleus with the use of different models for the charge distribution. In Appendix \ref{app:mu}, we recalculate this correction
using the exponential charge distribution to obtain results in agreement with \textcite{sgk2018, faustov2017}. 
Moreover, we note that this contribution is dominated by the mean square charge radius with the radiatively corrected
muon density at the nucleus. Therefore, the inelastic contribution is not expected to be significant,
and we estimate the relative uncertainties of our results in Eq.\ (\ref{E12}) at $10\%$.  

\section{Summary}
\label{section6}
If the nuclei were pointlike particles, the $2S-2P$ Lamb shift in light muonic atoms  would be sensitive to the hVP, 
as the muon $g-2$ is.  A theoretical understanding of nuclear structure at the relevant level of precision remains, 
despite recent steady progress, a challenging matter. As a result, the theoretical predictions for the energy spectra of muonic atoms
are currently a factor of $10$ to $\sim 100$ times less accurate than what would be obtained in the pointlike limit; 
therefore, the sensitivity to new physics in measurements in light muonic atoms is at present limited.  

The uncertainty of theoretical calculations are at present dominated by the hadronic and nuclear contributions rather than the QED terms, 
which can be obtained with much higher accuracy.
Focusing on such dominant contributions, it is useful to distinguish between single-nucleon and few-nucleons uncertainties.
The fact that the structure of the single nucleon is not well known is affecting terms such as 
$E_\mathrm{nucl1}$ in Eq.~\eqref{90} and generates the entire uncertainty of $E_\mathrm{NS}$ in $\mu$H. Presently such theoretical uncertainty for $\mu$H is of the order of the current experimental uncertainty. 
 Using only $\chi$PT makes it difficult to match the projected new measurements, which plan to achieve a factor of $5$ improvement in the empirical uncertainty. However, there are promising developments in the data-driven approach~\cite{tomalak2016}, 
and  lattice QCD calculations could achieve the needed precision in the future; see \cite{Fu:2022fgh} for the first calculation of the TPE contribution in $\mu$H. Furthermore, \citet{Hagelstein:2020awq} showed that the TPE contribution can be obtained using an alternative subtraction at $T_1(iQ,Q^2)$, which holds advantages for EFT and lattice QCD calculations. \\
The uncertainties related to the few-nucleon dynamics start at $\mu$D and move to heavier muonic atoms. 
They are due to the model dependence intrinsic to the parametrization of nuclear interactions. 
Calculations  have been performed using different nuclear potentials  to allow for an estimate of 
the associated model dependence~\cite{chen2013, dinur2016, bacca2018}. When the study is restricted to interactions developed within $\chi$EFT, an order-by-order analysis in the chiral expansion is necessary to estimate the uncertainty introduced in the truncation of the expansion. To date this has been achieved only for $\mu$D~\cite{hernandez2014, hernandez2017} and work is in progress for $\mu^3$He$^+$ and $\mu^4$He$^+$~\cite{muli2022}.  
Reducing such uncertainty is difficult and can be done only either by increasing the order of the $\chi$EFT expansion or by exploring other ways of fitting the $\chi$EFT low-energy constants at the present order.
Even if we were able to reduce these errors,  there are other sources of uncertainty, such as  corrections to Eq.~\eqref{90} including  higher-order polarizabilities and 
unknown corrections to  the nuclear electric dipole operator~\cite{Wienczek:2014mia,hernandez2019, bayesian}.

When the uncertainties stemming from single-nucleon and few-nucleon dynamics are compared for $\mu$D and  $\mu$He$^+$ the two are are found to be comparable in size even though the absolute contribution of terms stemming from the few-nucleon dynamics is larger; see Table 7 of~\cite{bacca2018}.
Finally, another important source of uncertainty  in  $\mu$He$^+$ ion is the unknown inelastic part of the three-photon exchange correction; see Sec.~\ref{n03}.

In view of these uncertainties, the comparison of the nuclear rms charge radii between muonic and electronic atoms would be interesting. 
Recent results using hydrogen spectroscopy \cite{H1, H2, H3, H4, H5}, although not yet conclusive, tend to be in agreement with
the $\mu$H value. The absolute determination of nuclear radii from the optical spectroscopy of normal atoms, other than   hydrogen,
has thus far not been successful. The only attempt from the measurement of the $2^3S - 2^3P$ transition in $^4$He \cite{pachucki_helium}, 
although in agreement with the $\mu^4$He$^+$ determination,  is much less accurate
due to the high complexity of QED effects in systems consisting of more than one electron. On the other side,
the optical spectroscopy of the one-electron He$^+$ ion has not yet been accomplished \cite{eikema, udem}.

The proton rms radius $r_p$ extracted from the $\mu$H Lamb shift (see Table~\ref{tab:recfns})  is by far the most accurate. 
Therefore, one can use this $r_p$ for the most accurate determination of the Rydberg constant from the $1S-2S$ hydrogen spectroscopy, 
for the most accurate determination of the rms deuteron 
charge radius $r_d$ from the H-D isotope shift in the $1S-2S$ transition, or recently for the most accurate determination of 
the electron-proton mass ratio from the precise spectroscopy of the HD$^+$ molecule.  The Lamb shift measurements in all other muonic atoms, 
although they do not lead to improvements in tests of fundamental physics, can give valuable information about
electromagnetic properties of nuclei.
Namely, we recall that the energy shift due to the finite nuclear size is proportional to the nuclear charge radius ($E_\mathrm{FNS}=\mathcal{C}r_C^2$). It turns out that the weighted isotope shift in muonic atoms $E_L({\rm D})/{\cal C}_\mathrm{D} - E_L({\rm H})/{\cal C}_\mathrm{H}$, 
where the corresponding coefficients ${\cal C}$ are given in Table~\ref{tab:recfns},
can be used for the determination of the difference of squared nuclear charge radii with
a higher precision than the  individual charge radii due to partial cancellations of uncertainties, resulting in
\begin{align}
r_d^2-r_p^2|_\mathrm{muonic} = 3.820\,0(7)_\mathrm{exp}(30)_\mathrm{theo}\;\mathrm{fm}^2\,.
\label{119}
\end{align}  
Equation \eqref{119} is in perfect agreement with the value obtained from the electronic H-D isotope shift in the $1S-2S$ transition \cite{HDiso},
resulting in a much more accurate determination \cite{pachucki2018},
\begin{align}
r_d^2-r_p^2|_\mathrm{electronic}= 3.820\,7(3)\;\mathrm{fm}^2\,.
\end{align} 
This indicates that we have a good understanding of the electromagnetic properties of the deuteron. 
An analogous comparison can be performed for the $^3$He-$^4$He isotope shift, for which we obtain
\begin{align}
\frac{E_L(^4\mathrm{He}^+)}{{\cal C}_{^4\mathrm{He}^+}} - \frac{E_L(^3\mathrm{He}^+)}{{\cal C}_{^3\mathrm{He}^+}} = 
0.258\,5(30)\,\mathrm{fm}^2 +r_\alpha^2 - r_h^2,
\end{align}
where we took advantage of the partial cancellation of uncertainties in the nuclear structure contribution $E_\mathrm{NS}$.
A recent measurement of the Lamb shift in $\mu^3$He$^+$ \cite{crema:2023} gave
\begin{align}
r_h^2 - r_\alpha^2|_\mathrm{muonic} = 1.063\,6(6)_\mathrm{exp}(30)_\mathrm{theo}\,\mathrm{fm}^2
\label{122}
\end{align}
The value \eqref{122} deviated by $3.6\,\sigma$ from from the recent measurement of the $^3$He - $^4$He isotope shift in $2^3S_1 - 2^1S_0$ transition \cite{vanderwerf:2023}
\begin{align}
r_h^2 - r_\alpha^2|_\mathrm{electronic} = 1.075\,7(15)\,\mathrm{fm}^2
\end{align}
We point out, however, that the other isotope shift measurements in  $2^3P_1 - 2^3S_1$ transition
have thus far not been conclusive, because they are in contradiction with each other \cite{he_iso_theory}; see also \citet{vanderwerf:2023}.
Therefore, we postpone drawing conclusions until measurements of the $2^3P_1 - 2^3S_1$ transition in $^{3,4}$He are confirmed.

Finally, our determination of the proton, the deuteron, and the $\alpha$ particle charge radii differs from the original ones given by \textcite{antognini2013}, \textcite{pohl2016}, and \textcite{krauth2021} (see Table~\ref{tab:recfns}), especially for the deuteron, while that for the helion charge radius \cite{crema:2023} is based on our calculations presented here.
The main reason for these differences is the neglect of the eVP$^{(1)}$ correction to the TPE and the inelastic 3PE in the original determination.
Moreover, the large uncertainties in $E_{\rm NS}$ indicate that a more accurate calculation of the electromagnetic nuclear structure of light nuclei 
is necessary to explore the great potential of the precision muonic atom spectroscopy.

\begin{acknowledgments}
We are grateful to S.~G.~Karshenboim for explanations regarding his work, 
and to O.~Tomalak for letting us know the result of his calculation of the recoil correction in TPE with the proton.
K.P.\ was supported by  the National Science Centre (Poland) Grant No. 2017/27/B/ST2/02459.
F.H.\ and V.L.\ were supported by the Deutsche Forschungsgemeinschaft (DFG) through the Emmy Noether Programme under Grant No.\ 449369623. 
S.S.L.M., S.B., and R.P.\ were supported through the Cluster of Excellence ``Precision Physics, Fundamental Interactions,
and Structure of Matter" (PRISMA$^+$ EXC 2118/1) funded by the DFG within the German Excellence Strategy
(Project ID No.\ 390831469). R.P.\ acknowledges support from the European Union’s Horizon 2020 research and innovation program through the STRONG-2020 project under Grant Agreement No.\ 824093.
\end{acknowledgments}

\appendix

\section{Subtraction of the point deuteron amplitude}
\label{app:subtraction}
In this appendix we specify the changes to the covariant dispersive formalism of \citet{carlson2014} that were
necessary to account for the subtraction of the point deuteron amplitude described in Sec.~\ref{n01}. As explained there, \citet{carlson2014} assumed that $g=0$ for the pointlike deuteron, whereas the definition of the latter as a Proca particle means that $g=1$. To compensate for this difference, one needs to modify Eq.~(16) of \citet{carlson2014}, which describes the elastic contribution of the TPE in terms of the elastic deuteron form factors as follows:
\begin{align}
E^\mathrm{el} & = \frac{m \alpha^2}{M(M^2-m^2)}\phi(0)^2
\int\limits_0^\infty\frac{dQ^2}{Q^2} 
\nonumber\\
&
 \times\left\{
\frac{2}{3}\left[G_M^2-1\right]
(1+\tau)\hat{\gamma}_1(\tau,\tau_l)-\frac{2}{3}(\tau-\tau_l)\frac{\gamma_1(\tau_l)}{\sqrt{\tau_l}}
 \right. \nonumber\\
& \qquad \left. 
-\left[\frac{G_C^2-1}{\tau}+\frac{2}{3}\left[G_M^2-1\right]+\frac{8}{9}\tau G_Q^2\right]
\hat{\gamma}_2(\tau,\tau_l)
\right.\nonumber\\
& \qquad +16M^2\frac{M-m}{Q}G_C'(0)
\bigg\},
\label{eq:contrib_elastic}
\end{align}
where $Q^2=-q^2$, $\tau=Q^2/(4M^2)$, $\tau_l=Q^2/(4m^2)$, and the weighting functions are defined as they were by \citet{carlson2014},
\begin{subequations}
\begin{align}
\hat{\gamma}_{1,2}(x,y) & = \frac{\gamma_{1,2}(x)}{\sqrt{x}}-\frac{\gamma_{1,2}(y)}{\sqrt{y}},\\
\gamma_1(x) & = (1-2x)\sqrt{1+x}+2x^{3/2},\\
\gamma_2(x) & = (1+x)^{3/2}-x^{3/2} -\frac{3}{2}\sqrt{x}.
\end{align}
\end{subequations}
Note the second term in the curly brackets in Eq.~\eqref{eq:contrib_elastic} that is generated by the nonpole part of the point deuteron amplitude, 
and that the Thomson term still needs to be subtracted from the nonpole amplitude as it was by \citet{carlson2014}.
The numerical effect of the extra subtraction terms in Eq.~\eqref{eq:contrib_elastic} on $E_\mathrm{TPE}(\mu\text{D},2S)$ does not depend on the elastic deuteron form factors
and turns out to be small, namely, $-0.000\,038$~meV.

\section{eVP$^{(1)}$ correction with TPE on single nucleons}
\label{AppendixEVP}

In this appendix we give an improved estimate for the eVP$^{(1)}$ correction  to the TPE between the muon and individual nucleons [cf.\ Eqs.~\eqref{91} and \eqref{103}],
\begin{align}
 \delta E_{\rm nucl1} =& -\frac{\pi}{3}\,m\,\alpha^2\phi^2(0)
  \Big[Z\,\delta[R_{F}^3(p)] + (A-Z)\,\delta[R_{F}^3(n)]\Bigr]\nonumber \\ &
 + 2\,\frac{\delta\phi(0)}{\phi(0)}\,E_{\rm nucl1}\,. \label{B1}
\end{align}
For the Born TPE with an eVP$^{(1)}$ insertion, we use the nucleon form factor parametrizations from \citet{Borah:2020gte}. 
For the nucleon polarizability contribution with an eVP$^{(1)}$ insertion, 
we use the leading-order $\chi$PT prediction plus the contribution of the $\Delta(1232)$ intermediate state 
(with the latter equal for $p$ and $n$); see \textcite{lensky2018,lensky2022a,Alarcon:2013cba}. 
Our total results for TPE with eVP$^{(1)}$ insertion are 
\begin{align}
\delta[R_{F}^3(p)] &= 0.053(10)\;\mathrm{fm}^3\,, \label{B2}\\
\delta[R_{F}^3(n)] &= 0.017(10)\;\mathrm{fm}^3\,. \label{B3}
\end{align}
The wave function correction was taken from Table 13 of \citet{sgk2021},
\begin{align}
2\,\frac{\delta\phi(0)}{\phi(0)} = \frac{\alpha}{\pi}\,\left\{
\begin{array}{ll}
1.404\,3 & \mathrm{for}\; \mu\mathrm{H}\\
1.452\,3 & \mathrm{for}\; \mu\mathrm{D}\\
2.181\,8 & \mathrm{for}\; \mu^3\mathrm{He}^+\\
2.192\,0 & \mathrm{for}\; \mu^4\mathrm{He^+}\,.
\end{array}\right.
\end{align}
Adding all up, we find the total eVP$^{(1)}$ correction to TPE with individual nucleons,
\begin{align} \label{B5}
\delta E_\mathrm{nucl1}(2S) = \left\{
\begin{array}{lll}
-0.000\,6(1) & \mathrm{meV} & \mathrm{for}\; \mu\mathrm{H}\\
-0.001\,0(2)\ &\mathrm{meV} & \mathrm{for}\; \mu\mathrm{D}\\
-0.016(2)   & \mathrm{meV} & \mathrm{for}\; \mu^3\mathrm{He}^+\\
-0.018(3)  & \mathrm{meV} & \mathrm{for}\; \mu^4\mathrm{He}^+\,.
\end{array}\right.
\end{align}

\section{eVP$^{(1)}$ correction with TPE on different nucleons}
\label{AppendixEVA}
This derivation is based on \citet{pachucki2015}. 
We now consider the muonic matrix element $P$
for the nonrelativistic two Coulomb exchange
\begin{equation}
P =\biggl\langle\phi\biggl|\frac{\alpha}{|\vec r-\vec r_a|}
\frac{1}{(H_0-E_0+E)} \frac{\alpha}{|\vec r-{\vec r_b}|}\biggr|\phi\biggr\rangle,
\label{C1}
\end{equation}
where $E$  is the nuclear excitation energy.
Using the on-mass-shell approximation and subtracting the point Coulomb exchange, Eq.~\eqref{C1} becomes
\begin{align}
P =&\, \alpha^2\,\phi^2(0)
\int\frac{d^3 k}{(2\,\pi)^3}\,\biggl(\frac{4\,\pi}{\vec k^{\,2}}\biggr)^2
\biggl(E+\frac{\vec k^{\,2}}{2\,\mu}\biggr)^{-1}
\nonumber \\ &\times \bigl(e^{i\,\vec k\cdot(\vec r_a- \vec r_b)}-1 \bigr).
\label{C2}
\end{align}
We now calculate the expansion coefficients in powers of $E$.
There are two characteristic integration regions: $|\vec k|\sim \sqrt{E\,m}$ and
$|\vec k|\sim m$. In the first integration region, where $|\vec k|$ is small, 
one performs an expansion of the exponent in powers of \mbox{$\vec k\cdot(\vec r_a - \vec r_b)$}. 
The leading quadratic term is the electric dipole contribution 
\begin{eqnarray}
P_\mathrm{low} =& 
\frac{4\,\pi}{3}\,\alpha^2\phi^2(0)\sqrt{\frac{2\,\mu}{E}}\,\vec r_a\cdot\vec r_b\,,
\label{C3}
\end{eqnarray}
which gives $E_\mathrm{pol}$ in Eq. (\ref{93}). 
In the second integration region, where $|\vec k|\sim m$ is large, one performs an expansion
in powers not exactly of $E$ but of the total nuclear energy $\tilde E$, 
\begin{equation}
\tilde E = E + \frac{\vec k^{\,2}}{2\,M}\,. \label{C4}
\end{equation} 
The first expansion term after integration over $\vec k$  is
\begin{equation}
P_\mathrm{high} = \frac{\pi}{3}\,m\,\alpha^2\,\phi^2(0)\,|\vec r_a-\vec r_b|^3 \label{C5}
\end{equation}
and the corresponding correction to the energy is $E_\mathrm{nucl2}$ in Eq.\ (\ref{92}).

To obtain the eVP$^{(1)}$ correction to $E_\mathrm{nucl2}$, we modify one of the Coulomb propagators in Eq.\ (\ref{C2}),
subtract the finite size with the leading polarizability
\begin{align}
\delta P =&\, \alpha^2\,\phi^2(0)
\int\frac{d^3 k}{(2\,\pi)^3}\,\biggl(\frac{4\,\pi}{\vec k^{\,2}}\biggr)^2\,\frac{2\,m}{\vec k^{\,2}}\,\bigl[-2\,\bar\omega^{(1)}(-\vec k^{\,2}/m_e^2)\bigr]
\nonumber \\ &\times \bigl\{e^{i\,\vec k\cdot(\vec r_a- \vec r_b)} - 1 +  \bigl[\vec k\cdot(\vec r_a - \vec r_b)]^2/2\bigr\},
\label{C6}
\end{align}
and use the large-$|\vec k|$ behavior of $\bar\omega^{(1)}$,
\begin{align}
\bar\omega^{(1)}(-\vec k^{\,2}/m_e^2) \approx \frac{\alpha}{3\,\pi}\,\left(5-\ln\frac{ \vec k^{\,2}}{m_e^2}\right)\,,
\end{align}
to obtain
\begin{equation}
\delta P = -\frac{4}{9}\,m\,\alpha^3\,\phi^2(0)\,|\vec r_a-\vec r_b|^3\,\Bigl[\frac{5}{12} + \ln(m_e\, |\vec r_a-\vec r_b |) +\gamma_E\Bigr]
\end{equation}
The corresponding correction to the energy is
\begin{align}
\delta E_\mathrm{nucl2} =&\ 2\,\frac{\delta\phi(0)}{\phi(0)}\, E_\mathrm{nucl2} -\sum_{a,b}\,\langle\delta P\rangle\nonumber \\
=&\  E_\mathrm{nucl2}\,\Bigl[ 2\,\frac{\delta\phi(0)}{\phi(0)} -\frac{\alpha}{\pi}\,\frac{4}{3}\,\Bigl(\frac{5}{12}+\ln(m_e\,r_L) +\gamma_E\Bigr)\Bigr]\,,
\end{align}
where $r_L$ is defined by
\begin{align}
\sum_{a,b}\,\langle|\vec r_a-\vec r_b|^3\,\ln|\vec r_a-\vec r_b|\rangle = \sum_{a,b}\,\langle|\vec r_a-\vec r_b|^3\rangle\,\ln r_L\,.
\end{align}
To estimate $\delta E_\mathrm{nucl2}$, we assume that $r_L\approx 2\,r_C$,
use the result of \citet{bacca2018} for $ \delta^{(1)}_{R3}=E_\mathrm{nucl2}(2S)  = -8.625$~meV and $-3.580$~meV for $\mu^3$He$^+$ and $\mu^4$He$^+$, 
respectively,  and obtain
\begin{align}
\label{C11}
\delta E_\mathrm{nucl2}(2S) =\left\{
\begin{array}{llll}
-0.140 \ (21) & \mathrm{meV} & \mathrm{for}& \mu^3\mathrm{He}^+\\
-0.060 \ (10) & \mathrm{meV} & \mathrm{for} & \mu^4\mathrm{He}^+\,,
\end{array}\right.
\end{align}
where the relative uncertainties of $15\%$ and $16\%$ for $\mu^3\text{He}^+$ and $\mu^4\text{He}^+$ result from the uncertainties of the adopted values of $E_\mathrm{nucl2}$ and $r_L$ ($2\%$ and $15\%$ for $\mu^3\text{He}^+$; $5\%$ and $15\%$ for $\mu^4\text{He}^+$) summed in quadrature.

\section{eVP$^{(1)}$ correction with the leading polarizability}
\label{AppendixEpol}

The derivation is based on the work of \citet{kalinowski2019}.
We now consider the leading polarizability in Eq.\ (\ref{93}) that comes from the TPE
\begin{align}
E_{\rm pol} =&\   
-\frac{4\,\pi\,\alpha^2}{3}\,\phi^2(0)\int_{E_T}dE\, |\langle\phi_N|\,\vec{\!d}\,|E\rangle|^2\, \sqrt{\frac{2\,\mu}{E}}\,.
\end{align}
The eVP$^{(1)}$ correction modifies one of the photon propagators, which leads to a complicated expression.
When the smallness of the parameter $m_e/\sqrt{E\,\mu}$ is taken into account, the first two terms in its expansion are
\begin{align}
\delta E_{\rm pol} =& -\frac{8\,\alpha^3}{9}\,\phi^2(0)\int_{E_T} d E\,\lvert\braket{\phi_N|\,\vec{\!d}\,|E}\rvert^2\sqrt{\frac{2\,\mu}{E}} \nonumber \\
&\times\left[\ln\left(\frac{2\,\mu\,E}{m_e^2}\right)-\frac{5}{3}+\frac{3\,\pi}{2}\,\sqrt{\frac{m_e^2}{2\,\mu\,E}}\right]\nn
\\ & + 2\,\frac{\delta\phi(0)}{\phi(0)}\,E_{\rm pol} \,,
\end{align}
where the last term comes from the eVP$^{(1)}$ correction to the wave function. 
\citet{kalinowski2019} obtained for $\mu$D the value $\delta E_{\rm pol}(2S) =  -0.026\,5(3)$ meV.
Using the wave function correction from Eq.\ (\ref{B5})
and performing calculations with the Argonne $v18$ potential~\cite{av18},   we obtain 
$\delta E_{\rm pol}(2S) =  -0.026\,3$ meV, which is in agreement with the aforementioned result.
Following the calculation by \citet{bacca2018} of the leading polarizability with including the Urbana IX three-body force,
we obtain \cite{muli2022}
\begin{align}
\label{D3}
\delta E_{\rm pol}(2S) = \left\{
\begin{array}{lll}
-0.110 \ (11) & \mathrm{meV} & \mathrm{for}\; \mu^3\mathrm{He}^+\\
-0.080 \ (6) & \mathrm{meV} & \mathrm{for}\; \mu^4\mathrm{He}^+\,,
\end{array}\right.
\end{align}
where the relative uncertainty for $\mu^4\text{He}^+$ is $7 \%$, distributed as follows: $3 \%$ from the missing multipoles, $5 \%$ from the nuclear model, and $4 \%$ from the numerical evaluation. For $\mu^3\text{He}^+$, the relative uncertainty is $10 \%$, which comes mostly from the missing multipoles ($9 \%$), with smaller additional uncertainties of $2\%$ from the nuclear model and $4 \%$ from the numerical evaluation. 

\section{$\mu$SE$^{(1)}$ + $\mu$VP$^{(1)}$ correction with the elastic TPE}
\label{app:mu}
This derivation is based on the work of \citet{pachucki1993}.
 In the limit of an infinite mass nucleus, the elastic TPE with $\mu$SE$^{(1)}$ + $\mu$VP$^{(1)}$ 
 corrections and point-nucleus subtraction reads
\begin{align}
\delta E_\mathrm{TPE} =&\ \frac{\alpha}{\pi}\,\frac{\phi^2(0)}{m^2}\,\int\frac{d^3p}{(2\,\pi)^3}\,\frac{(4\,\pi\,\alpha)^2}{p^4}\,f(p^2)\,\left[\rho(p^2)^2-1\right]\,,
\end{align}
where $\rho$ is the nuclear charge form factor,
$f(p^2)$ is the radiatively corrected muon line at the momentum exchange $p^0=0$ [see Eq.\ (\ref{104})],
\begin{align}
\frac{\alpha}{\pi}\,f(p^2) = \frac{1}{2}\,t^{\rm rad}_{00} +\,\frac{4}{p^2}\,\bar\omega^{(1)}(p^2)\,,
\end{align}
and all momenta are in the muon mass units.
$f(p^2)$ was calculated by \citet{pachucki1993} using dispersion relations,
\begin{align}
f(p^2) =&\ -\int_0^\infty d(q^2)\,\frac{f^A(q^2)}{q^2+p^2}\,, \label{A3}
\end{align}
with
\begin{align}
f^A(q^2) =&\  \frac{q^2}{4} \bigg(\frac{1}{1+q^2}-J^A\bigg) +\bigg(\frac{4}{q^2}+1\bigg) (J^A-1) \nonumber \\ &\hspace*{-7ex}
+ \Theta(q-2)\bigg(\frac{4}{q^2}+1\bigg) \bigg(\frac{1}{q^2\,\sqrt{1-\frac{4}{q^2}}}+\sqrt{1-\frac{4}{q^2}}\bigg)\nonumber \\ & \hspace*{-7ex}
+ \frac{4}{3}\, \Theta(q-2) \sqrt{1-\frac{4}{q^2}}\,\frac{1}{q^2}\, \bigg(1+\frac{2}{q^2}\bigg)\,,
\end{align}
where the last term comes from $\mu$VP and
\begin{align}
J^A = \frac{1}{q}\,\bigg[\arctan(q)-\Theta(q-2)\, \arccos\bigg(\frac{2}{q}\bigg)\bigg]\,.
\end{align}
Using Eq.\ (\ref{A3}), one can transform $\delta E_\mathrm{TPE}$ into
\begin{align}
\delta E_\mathrm{TPE} = \frac{\alpha}{\pi}\,\frac{\phi^2(0)}{m^2}\,\frac{(4\,\pi\,\alpha)^2}{2\,\pi^2}\,\int_0^\infty d(q^2)\,f^A(q^2)\,g(q)
\end{align}
where
\begin{align}
g(q) =   \int_0^\infty dp\,\frac{1}{p^2\,(q^2+p^2)}\,[1-\rho(p)^2]\,.
\end{align}
When one assumes a dipole parametrization of the electric form factor
\begin{align}
\rho(p^2) =&\ \frac{\Lambda^4}{(\Lambda^2+p^2)^2}\,,
\end{align}
$g(q)$ becomes
\begin{align}
g(q) =&\ \frac{\pi}{16\,q}\,h(q)\,,
\end{align}
where
\begin{align}
h(q) =&\  \frac{\Lambda^2}{(\Lambda+q)^4}+\frac{4 \Lambda}{(\Lambda+q)^3}+\frac{19}{2 (\Lambda+q)^2}+\frac{35}{2 \Lambda(\Lambda+q)}\,.
\end{align}
Finally, the correction to the energy
\begin{align}
\delta E_\mathrm{TPE} =&\ \alpha\,(Z\,\alpha)^2\,\frac{\phi^2(0)}{m^2}\,\int_0^\infty dq\, h(q)\, f^A(q^2)
\end{align}
is integrated numerically with $\Lambda = 2\,\sqrt{3}\,\lbar_\mu/r_C$, $r_p = 0.841$ fm, $r_d = 2.127$ fm, $r_h = 1.969$ fm, and $r_\alpha = 1.678$ fm,
to obtain 
\begin{align}
\label{E12}
\delta E_\mathrm{TPE}(2S) =\left\{
\begin{array}{lll}
-0.000\,4  & \mathrm{meV}\; \mathrm{for}& \mu\mathrm{H}\\
-0.002\,6(3)  & \mathrm{meV}\;\mathrm{for}& \mu\mathrm{D}\\
-0.077(8)  & \mathrm{meV} \; \mathrm{for}& \mu^3\mathrm{He}\\
-0.059(6)   & \mathrm{meV} \; \mathrm{for}& \mu^3\mathrm{He}\,,
\end{array}\right.
\end{align}
where we have assumed a 10\% uncertainty due to the elastic approximation.
These results are presented in Table~\ref{tab:recfns}.
%, where those for $\mu$D and $\mu$He$^+$ are divided by 2, as explained in subsection~\ref{n05}.
%\bibliography{muonic_bib}
%

\end{document}